\documentclass[a4paper,11pt]{article}
\pdfoutput=1
\usepackage{jcappub}
\usepackage{bbold}
\usepackage[x11names]{xcolor}
\usepackage{colortbl,hhline}
\usepackage[english]{babel}
\usepackage[utf8]{inputenc}
\usepackage{amsmath}
\usepackage{color}
\usepackage{amsfonts}
\usepackage{graphicx}
\usepackage{amssymb}
\usepackage{eufrak}
\usepackage{etoolbox}
\usepackage{amsmath}
\usepackage{empheq}
\usepackage[makeroom]{cancel}
\usepackage[most]{tcolorbox}
\usepackage{float}              % Activate [H] option to place figure HERE
\usepackage{subcaption}
\usepackage[normalem]{ulem}
\usepackage{xspace}
 \usepackage[export]{adjustbox}
\newtcbox{\mymath}[1][]{%
    nobeforeafter, math upper, tcbox raise base,
    enhanced, colframe=yellow!30!black,
    colback=yellow!30, boxrule=1pt,
    #1}
\def\be{\begin{equation}}
\def\ee{\end{equation}}
\def\bea{\begin{eqnarray}}
\def\eea{\end{eqnarray}}
\def\bean{\begin{eqnarray*}}
\def\eean{\end{eqnarray*}}

\def \dd {\partial}

\newcommand{\ie}{\emph{i.e.}\xspace}

\newcommand{\HH}{\mathcal H}

\definecolor{bittersweet}{rgb}{1.0, 0.44, 0.37}

\newcommand{\zetaf}{\zeta}

\newcommand {\gev}{\textit{gevolution}\xspace}
\newcommand {\kev}{{$k$-evolution}\xspace}

\usepackage{scalerel,tikz}
\usetikzlibrary{svg.path}
\definecolor{orcidlogocol}{HTML}{A6CE39}
\tikzset{orcidlogo/.pic={
 \fill[orcidlogocol] svg{M256,128c0,70.7-57.3,128-128,128C57.3,256,0,198.7,0,128C0,57.3,57.3,0,128,0C198.7,0,256,57.3,256,128z};
 \fill[white] svg{M86.3,186.2H70.9V79.1h15.4v48.4V186.2z}
 svg{M108.9,79.1h41.6c39.6,0,57,28.3,57,53.6c0,27.5-21.5,53.6-56.8,53.6h-41.8V79.1z M124.3,172.4h24.5c34.9,0,42.9-26.5,42.9-39.7c0-21.5-13.7-39.7-43.7-39.7h-23.7V172.4z}
 svg{M88.7,56.8c0,5.5-4.5,10.1-10.1,10.1c-5.6,0-10.1-4.6-10.1-10.1c0-5.6,4.5-10.1,10.1-10.1C84.2,46.7,88.7,51.3,88.7,56.8z};
}}
\newcommand\orcidicon[1]{\href{https://orcid.org/#1}{\mbox{\scalerel*{
\begin{tikzpicture}[yscale=-1,transform shape]
\pic{orcidlogo};
\end{tikzpicture}
}{|}}}}

\title{A new instability in clustering dark energy?}
\author[a,c]{Farbod Hassani~\orcidicon{0000-0003-2640-4460},}
\author[b]{Julian Adamek~\orcidicon{0000-0002-0723-6740},}
\author[c]{Martin Kunz~\orcidicon{0000-0002-3052-7394},}
\author[d,c]{Pan Shi}
\author[c]{and Peter Wittwer}
\affiliation[a]{
Institute of Theoretical Astrophysics, University of Oslo, P.O. Box 1029 Blindern, N-0315, Oslo, Norway
}
\affiliation[b]{
Institute for Computational Science, Universit\"at Z\"urich, Winterthurerstrasse 190, 8057 Z\"urich, Switzerland
}
\affiliation[c]{
 D\'epartement de Physique Th\'eorique and CAP, Universit\'e de Gen\`eve,
24 quai Ernest-Ansermet, CH-1211 Gen\`eve 4, Switzerland
}
\affiliation[d]{
School of Mathematics, Renmin University of China, Beijing 100872, People's Republic of China
}

\emailAdd{farbod.hassani@astro.uio.no} 
\emailAdd{julian.adamek@uzh.ch}  
\emailAdd{martin.kunz@unige.ch} 
\abstract{
In this paper, we study the effective
field theory (EFT) of dark energy for the $k$-essence model beyond linear order.
Using particle-mesh $N$-body simulations that consistently solve the dark energy evolution on a grid, we find that the next-to-leading order in the EFT expansion, which comprises the terms of the equations of motion that are quadratic in the field variables, gives rise to a new instability in the regime of low speed of sound (high Mach number). We rule out the possibility of a numerical artefact by considering simplified cases in spherically  and plane symmetric situations analytically. If the speed of sound vanishes exactly, the non-linear instability makes the evolution singular in finite time, signalling a breakdown of the EFT framework. The case of finite (but small) speed of sound is subtle, and the local singularity could be replaced by some other type of behaviour with strong non-linearities. While an ultraviolet completion may cure the problem in principle, there is no reason why this should be the case in general. As a result, for a large range of the effective speed of sound $c_s$, a linear treatment is not adequate.} 
\begin{document}
\maketitle

%%%%%%%%%%%%%%%%%
%%%INTRODUCTION
%%%%%%%%%%%%%%%%%
\section{Introduction}
\label{sec:intro}
In the near future, {cosmology} will benefit from numerous high precision observations \cite{Amendola:2016saw, Santos:2015gra, Aghamousa:2016zmz}, probing the Universe at different epochs, from the very early times when cosmic microwave background radiation (CMB) photons started to propagate and the Universe was around 400,000 years old to today, when the Universe is 13.8 billion years old and has entered a phase of accelerating expansion. One of the main goals of ongoing and future cosmological surveys is to elucidate the physical mechanism behind the late-time accelerated expansion of the Universe that has now been established by several independent observations \cite{Ade:2015xua, 2018ApJ...859..101S, 2017MNRAS.470.2617A}.

Over the past years, a wide range of theories has been developed by cosmologists and particle physicists with the aim to address the question of the accelerated expansion of the Universe, either by modifying the theory of gravity or by considering an additional fluid component with a negative pressure usually called dark energy (DE) \citep{Koyama:2018som, Clifton:2011jh, Joyce:2016vqv}.  Among these theories, the effective field theory (EFT) of dark energy \cite{Gubitosi:2012hu, Gleyzes:2014rba, Bellini:2014fua} has become quite popular since it allows to describe the dark energy phenomenology occurring at low energies with a reasonable number of free parameters in a generic way. In principle, the free parameters in the effective theory are connected to (\ie\ determined by) a fundamental theory at high energy and respect the low-energy symmetries \cite{Cheung:2007st}. In practice however, they can be considered as parameters to be measured in low-energy experiments, similar to the moduli describing an elastic material in the context of material science.

In this paper, we focus on the subset of EFT of DE models with only two free parameters, $\alpha_K$ (or equivalently $c_s^2=\delta p/\delta\rho$) at the perturbation level and $w=p/\rho$ at the background level, which is equivalent to the well-known theory of $k$-essence. The $k$-essence theory was first proposed in 2000 \cite{ArmendarizPicon:2000ah, ArmendarizPicon:2000dh, Vikman:2007sj} to naturally, without any fine-tuning, explain the accelerated expansion of the Universe. At linear level $k$-essence has been explored well and is considered a viable theory for the late-time accelerated expansion. Like the cosmological constant, which is reached in the limit $w\rightarrow-1$, this theory can explain all cosmological observations to date. However, by increasing the precision of the observations, we can hope to place much more stringent constraints on the space of theories.  In the near future this will require a good understanding of the behaviour at non-linear scales which requires developing proper $N$-body simulations \cite{ Li:2018qku, Baldi:2008ay, Hassani:2020rxd, Baldi:2012ua, Barreira:2013eea, Llinares:2013jza}. To capture the non-linear behaviour of $k$-essence we developed the code
$k$-evolution based on the relativistic $N$-body code \gev \cite{Adamek:2015eda, Adamek:2016zes, Adamek:2017uiq}. In previous studies \cite{Hassani:2019wed,Hassani:2020buk, Hassani:2020agf, Hansen:2019juz} we maintained linearity of the $k$-essence field equations and studied the evolution when coupled to a non-linear $N$-body system.
Here, we use the equations derived in \cite{Hassani:2019lmy} for the non-linear evolution of $k$-essence as an effective field theory, parametrised with the equation of state $w$ and the speed of sound $c_s$. The free parameters appearing in the field picture, e.g., $\alpha_K$, can be interpreted when writing the theory in the fluid picture. In Appendix A of \cite{Hassani:2019lmy}, we showed that the fluid description and the field picture are equivalent and 
one can easily change the picture by well-defined transformations.
 
In this paper, we show that the EFT of DE for $k$-essence, in the limit of low speed of sound, suffers from a new instability triggered by one of the non-linear terms in the EFT expansion. In Sec.~\ref{Sec:1_eq} we discuss the equations that describe the model.  In Sec.~\ref{sec_3_numerics} we present the numerical results for 3+1 D in the cosmological context where we solve the full 3+1 D partial differential equation for the $k$-essence scalar field numerically, using the EFT framework.
We show that for low speed of sound, the numerical solution to this partial differential equation (PDE) blows up at some time before the current age of the Universe. 
 In Sec.~\ref{sec:one_d} we study a simplified PDE, using either planar or spherical symmetry to reduce the dimensionality to 1+1 D. We show analytically that the instability is present and therefore expected to appear in the full 3+1 D case.
 We also comment on how the solution becomes singular and when the solution ceases to exist, and we show how increasing the speed of sound could stabilise the system. In the final section, we conclude with a short discussion of the results.

%%%%%%%%%%%%%%%%%
%%%Section :  Field equations
%%%%%%%%%%%%%%%%%

\section{Field equations} \label{Sec:1_eq}
In this section we write down the equations for the $k$-essence scalar field para\-me\-trised with $w$ and $c_s^2$, expanded around the background employing the weak-field expansion. The equations of motion as well as the stress energy tensor for clustering DE are obtained and discussed in detail in \cite{Hassani:2019lmy}.
There we showed the results for clustering DE where we only keep linear terms in the DE scalar field equations. 
In order to study the evolution of perturbations %in the Universe
we use the Friedmann-Lema\^itre-Robertson-Walker (FLRW) metric in the conformal Poisson gauge,
\be
ds^2 = a^2(\tau) \Big[ - e^{2 \Psi} d\tau^2 -2 B_i dx^i d\tau  + \big( e^{-2 \Phi} \delta_{ij} + h_{ij}\big)  dx^i dx^j \Big],
\ee
where $\Psi$ and $\Phi$ are the temporal and spatial scalar perturbations of the metric and correspond to the Bardeen potentials, $B_i$ is the transverse gravitomagnetic vector perturbation with two degrees of freedom, and $h_{ij}$ is the traceless transverse tensor perturbation with two degrees of freedom.
In this gauge the PDE  for the $k$-essence scalar field, including the non-linear corrections in the weak-field regime, reads 
\begin{equation}
 \partial_\tau \pi = \zetaf - \HH \pi + \Psi \;, \label{zeta_eq1}
\end{equation}
\begin{multline}
\label{zeta_eq2}
 \partial_\tau \zetaf =  3 w  \HH \zetaf -  3 c_s^2 \left(\HH^2 \pi - \HH \Psi -  \HH' \pi - \partial_\tau \Phi \right) +c_s^2 {\nabla}^2 \pi  \\
- \Big(\vec  \nabla \left[2 (c_s^2-1)  \zetaf +c_s^2    \Phi-    \Psi \right] \Big) \cdot \vec  \nabla \pi - \left[ (c_s^2-1)\zetaf    + c_s^2 \Phi - c_s^2  \Psi \right] \nabla^2 \pi \\
-  \frac{\HH}{2} \left[ (2+ 3 w +c_s^2 ) (\vec \nabla \pi)^2 +  6 c_s^2 (1+ w) \pi  \nabla^2 \pi  \right]     +\frac{c_s^2 -1 }{2} {\vec \nabla \cdot} \left((\vec \nabla \pi)^2  {\vec \nabla} \pi\right) \;. 
\end{multline}
In this equation $\pi$ is the DE scalar field and $\zeta$ is an auxiliary field written in terms of the scalar field $\pi$, its time derivative $\partial_\tau \pi$ with respect to conformal time and the gravitational potential $\Psi$. Moreover, $(\vec \nabla \pi)^2 \equiv \vec \nabla \pi \cdot  \vec \nabla \pi$, $\vec \nabla$ is the spatial gradient using partial derivatives and $\nabla^2$ is the corresponding Laplace operator.
As is typical for a weak-field expansion inside the horizon we keep any higher-order terms only if they contain at least two spatial derivatives for each power of a perturbation variable beyond the first order. A simple heuristic argument comes from observing that $(\vec\nabla\Psi)^2 /\mathcal{H}^2 \sim v^2 \sim \Psi$ and $\nabla^2\Phi /\mathcal{H}^2 \sim \delta \sim 1$, which implies that each spatial derivative effectively counts as $-1/2$ order. More details can be found in \cite{Hassani:2019lmy}.

As discussed in Appendix A of \cite{Hassani:2019lmy}, Eq.~\eqref{zeta_eq2} is equivalent to the continuity and the Euler equations,
\be
\label{conteq}
 \partial_\tau \delta = -(1+w) \big( \theta- 3 \partial_\tau \Phi \big) -3 \mathcal{H}  \bigg(\frac{\delta p} {\delta \rho} -w\bigg) \, \delta + 3  \partial_\tau\Phi  \bigg( 1+ \frac{\delta p} {\delta \rho}  \bigg) \, \delta+ \frac{1+w}{\rho} v^i \nabla_i  \big(3\Phi - \Psi \big) \;,
 \ee
\begin{multline}
\partial_\tau\theta+   (3w-1) \mathcal{H}  \, \theta +  \nabla^2 (\Psi + \sigma)+ \frac{\nabla^2 \delta P}{\rho(1+w)}  - ( 5 \partial_\tau\Phi +   \partial_\tau\Psi) \theta +  \frac{\nabla^2 \Psi }{1+w}\bigg(1+  \frac{ \delta P } {\delta \rho}  \bigg) \, \delta
\\ 
- \frac{  \nabla_i  \Sigma^{i j}} {\rho (1+w)} \nabla_j ( {3  \Phi }  -   \Psi  )=0 \,,
\end{multline}
where $\Sigma^{ij} = T^{ij} - \delta^{ij} T_k^k/3$ is the anisotropic stress tensor, $\delta^{ij}$ is the Kronecker delta, $ \nabla_i \equiv \partial_i$ denotes the partial derivative and we have used the following definitions,
\be
\label{definitions}
\delta \doteq \frac{\delta \rho}{\rho} \;, \qquad
\theta \doteq e^{-2(\Phi + \Psi)} \nabla_i v^i \;,  \qquad \sigma \doteq \frac{\nabla^{-2} \nabla_i \nabla_j \Sigma^{i j} }{\rho+p} \;,
\ee
and where $\nabla^{-2} $ is the inverse Laplace operator. The effective fluid {variables} read as follows in terms of the field {variables},
 \be
\label{SEfullfluids}
\begin{split}
\delta \rho & =  - \frac{\rho+p}{c_s^2} \bigg[ 3c_s^2 \HH \pi -\zetaf - \frac{2c_s^2 - 1}{2} (\vec{\nabla} \pi)^2   \bigg ] \;, \\
\delta p &  = -  \, (\rho+p)  \bigg[ 3 w  \HH \pi -\zetaf +  \frac16 (\vec{\nabla} \pi)^2   \bigg]  \;, \\
v^i & = - e^{2 (\Phi + \Psi)} \bigg[1- \frac{1}{c_s^2}   \, \Big( 3 c_s^2 (1+w) \mathcal{H} \pi -\zetaf + c_s^2 \Psi \Big) + \frac{c_s^2 - 1}{2 c_s^2}   (\vec{\nabla} \pi)^2  \bigg ] \delta^{ij} \nabla_j \pi \;, \\
\Sigma^{i j} & = (\rho+p) \left[ \delta ^{jk} \delta ^{il} \nabla_k \pi \nabla_l \pi - \frac13 ({\vec{\nabla}} \pi)^2 \delta^{i j} \right] \;.
\end{split}
\ee
However, in the implementation chosen for \kev, we solve the equations written in the field language and we solve a second-order PDE to update the scalar field $\pi$ and $\zeta$. Numerical results from \kev for the full non-linear PDE show that there exists a critical speed of sound $c_s^\ast$, such that for speed of sound $c_s$ smaller than $c_s^\ast$ the solution of the PDE becomes singular in finite time. In the following section we show numerical results for examples of small and large speeds of sound where the solution is, respectively, singular and regular, and afterwards we justify the numerical results by studying the equations in simpler setups with spatial symmetries that allow a corresponding reduction of the dimensionality of the problem.

 %%%%%%%%%%%%%%%%%
%%%Section :  Simulations
%%%%%%%%%%%%%%%%%

 \section{Results from cosmological simulations} \label{sec_3_numerics}
In this section we show the numerical results from {a first set of cosmological simulations with} \kev\ that include non-linear terms in the $k$-essence field equations. For simplicity we start with standard linear perturbations in matter that are derived from the $\Lambda$CDM model and set the two additional fields $\pi$ and $\zeta$ to zero initially. At very high redshift their contribution to the energy density is negligible and hence the matter solution is indeed the one of $\Lambda$CDM. The solution of $\pi$ and $\zeta$ will contain a decaying mode due to the way we set the initial conditions, but this will have no relevant effect on the final evolution if our initial redshift was chosen high enough. 

\begin{figure*}
\begin{minipage}{\textwidth}
%\centering
\vspace*{-2cm}
 {\hspace*{0cm}{\includegraphics[scale=0.38, trim=6cm 5cm 5cm 0cm, clip]{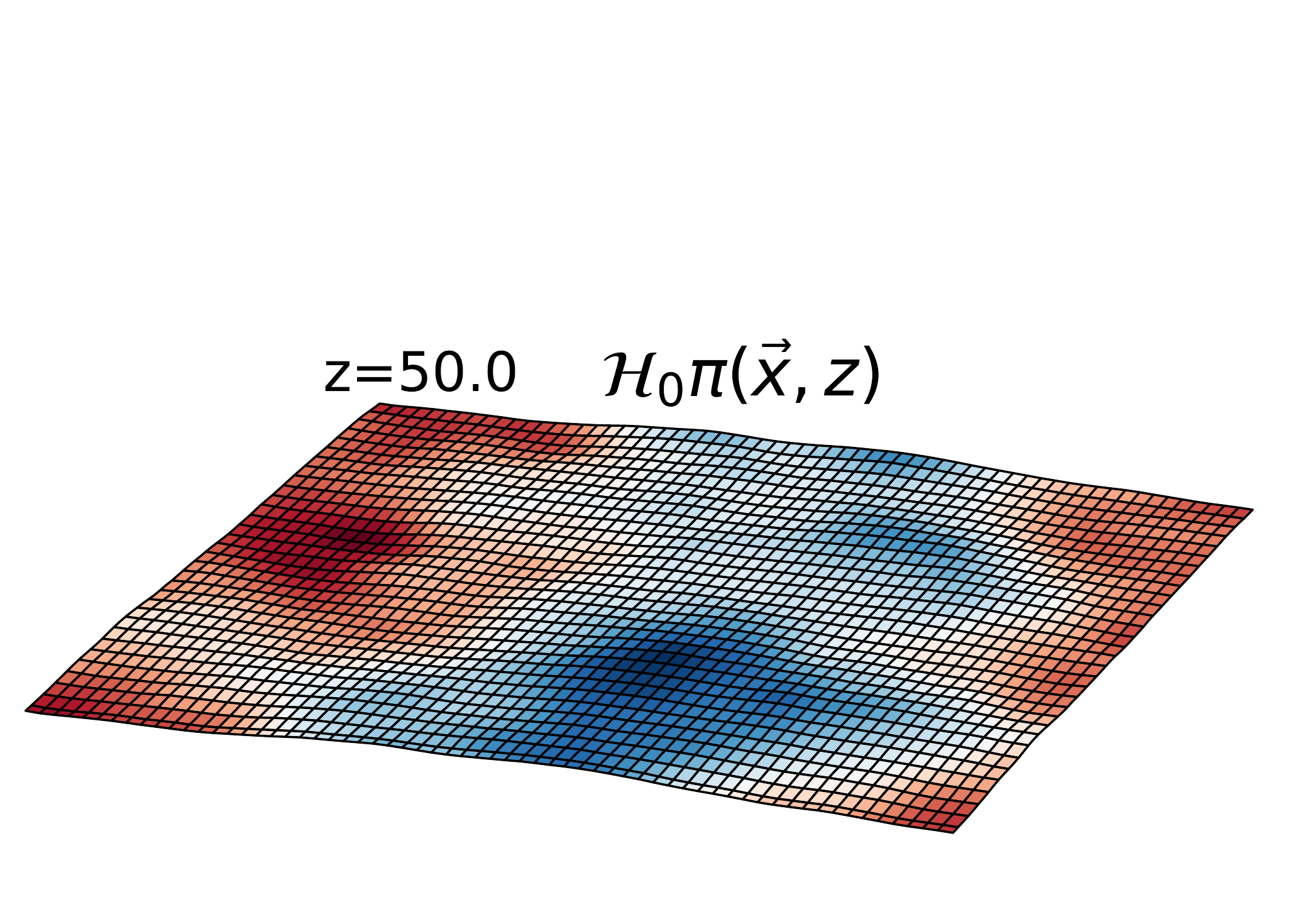} }}\hfill
 {{\includegraphics[scale=0.15, trim=-1cm 6.5cm 0cm 0cm]{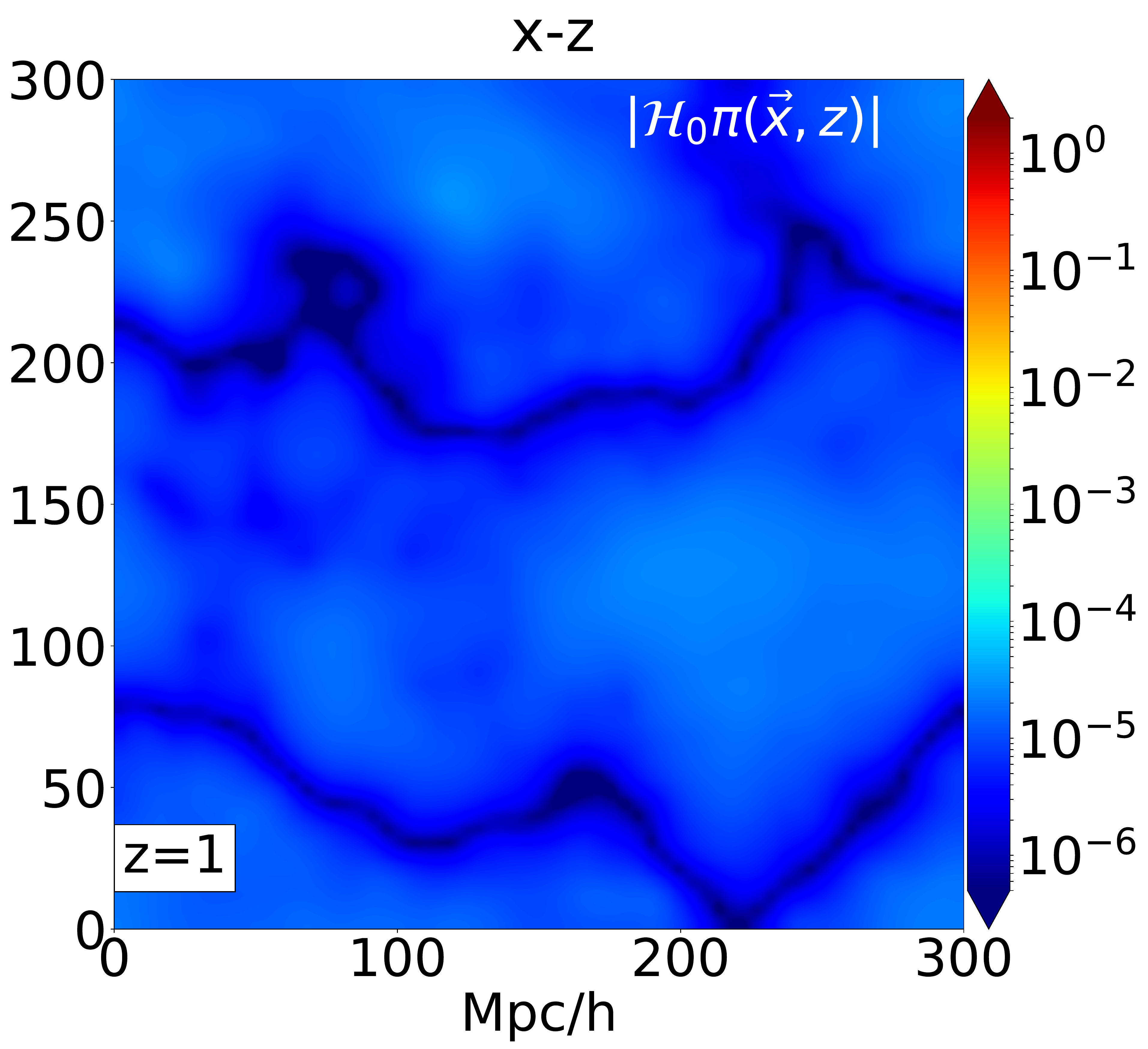} }}
  %%%%%%%%%%%%%%%%%%%%%%
 {\hspace*{0cm}{\includegraphics[scale=0.38, trim=6cm 4cm 5cm 2cm, clip]{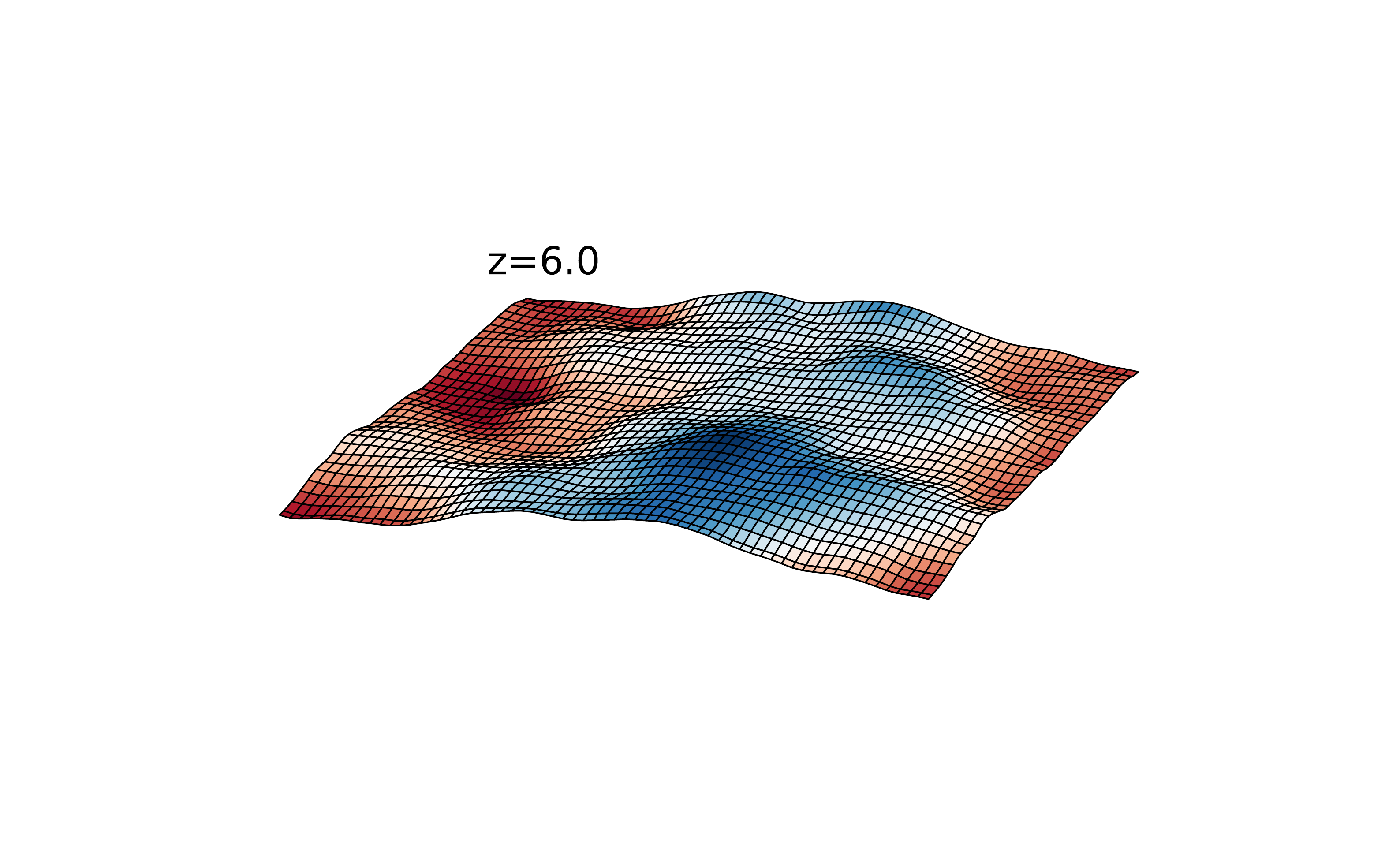} }}\hfill
 {{\includegraphics[scale=0.15, trim=-1cm 5cm 0 0cm]{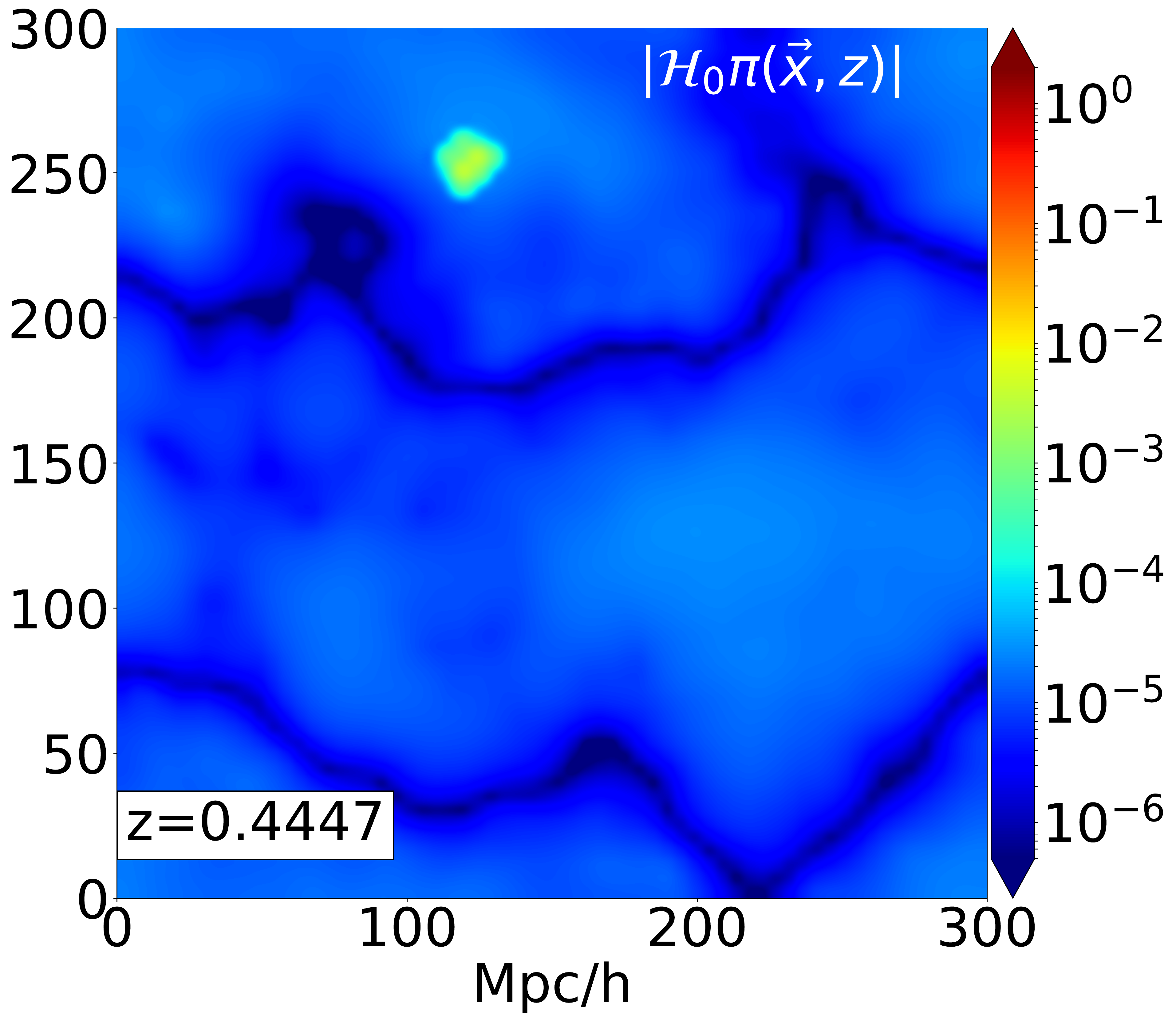} }}
  %%%%%%%%%%%%%%%%%%%%%%
  {\hspace*{0cm}{\includegraphics[scale=0.38, trim=6cm 4cm 5cm 3cm, clip]{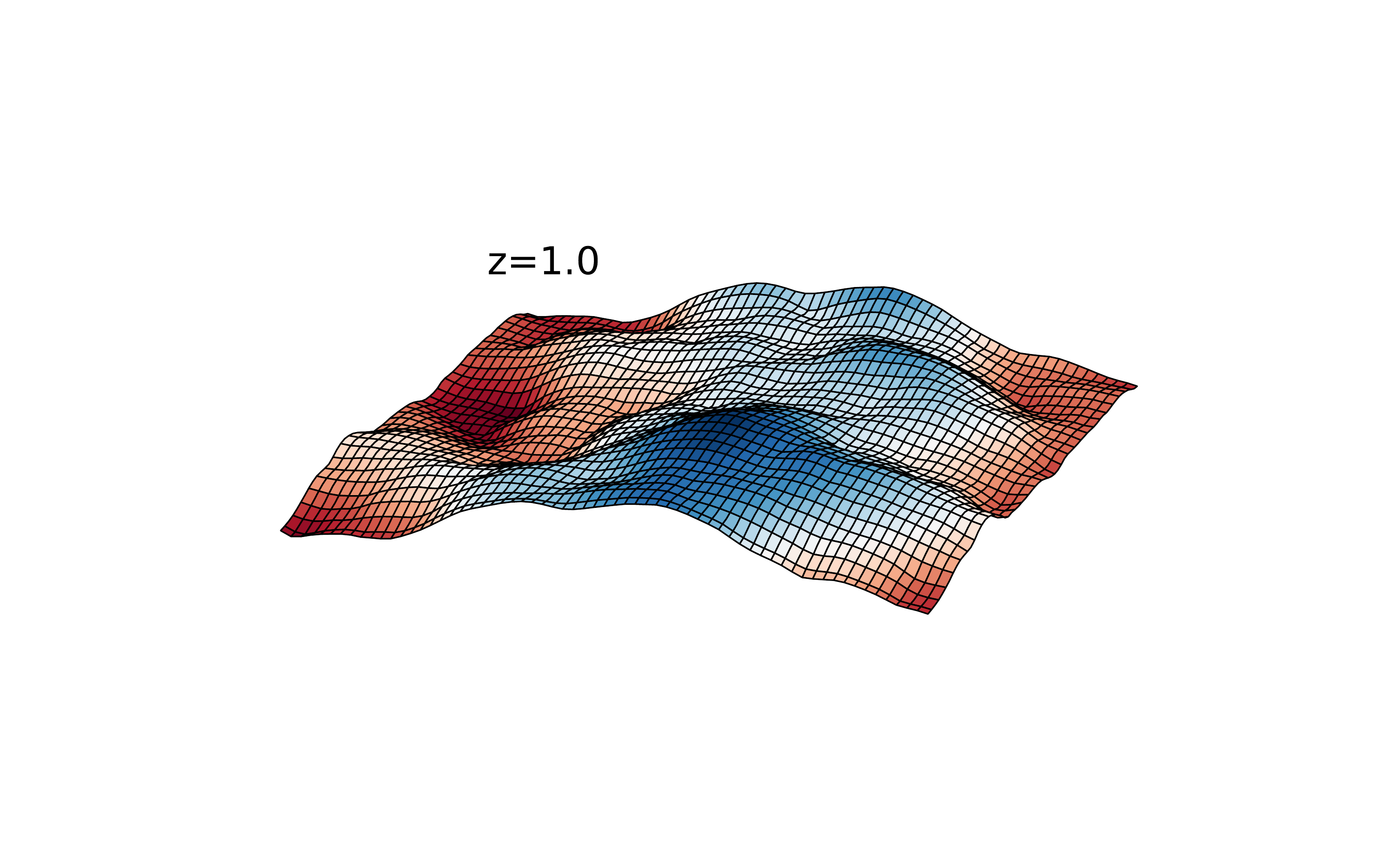} }}\hfill 
 {{\includegraphics[scale=0.15, trim=-1cm 6cm 0 -1cm]{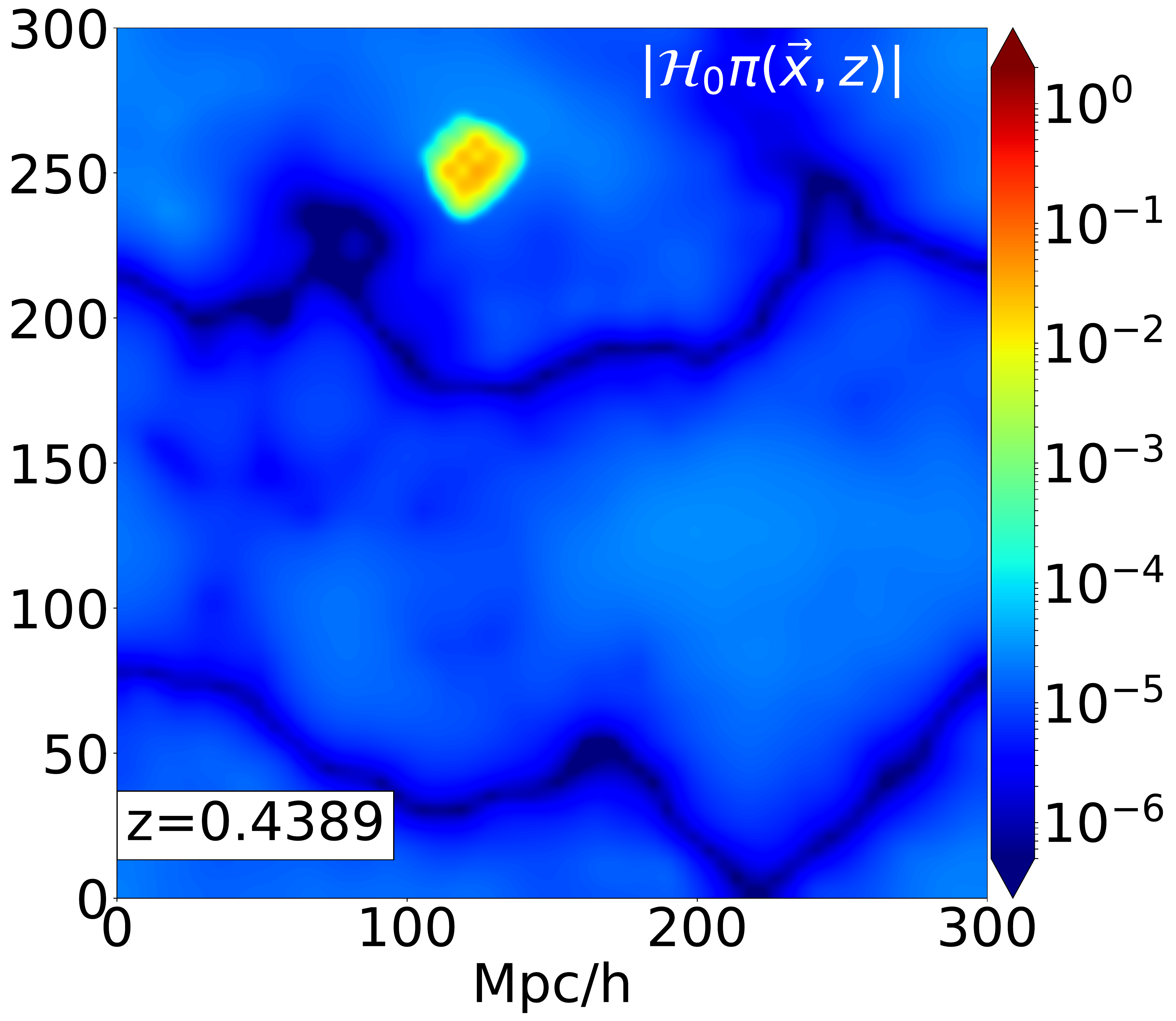} }}
  %%%%%%%%%%%%%%%%%%%%%%
 {\hspace*{0cm}{\includegraphics[scale=0.38, trim=6cm 2cm 5cm 3cm, clip]{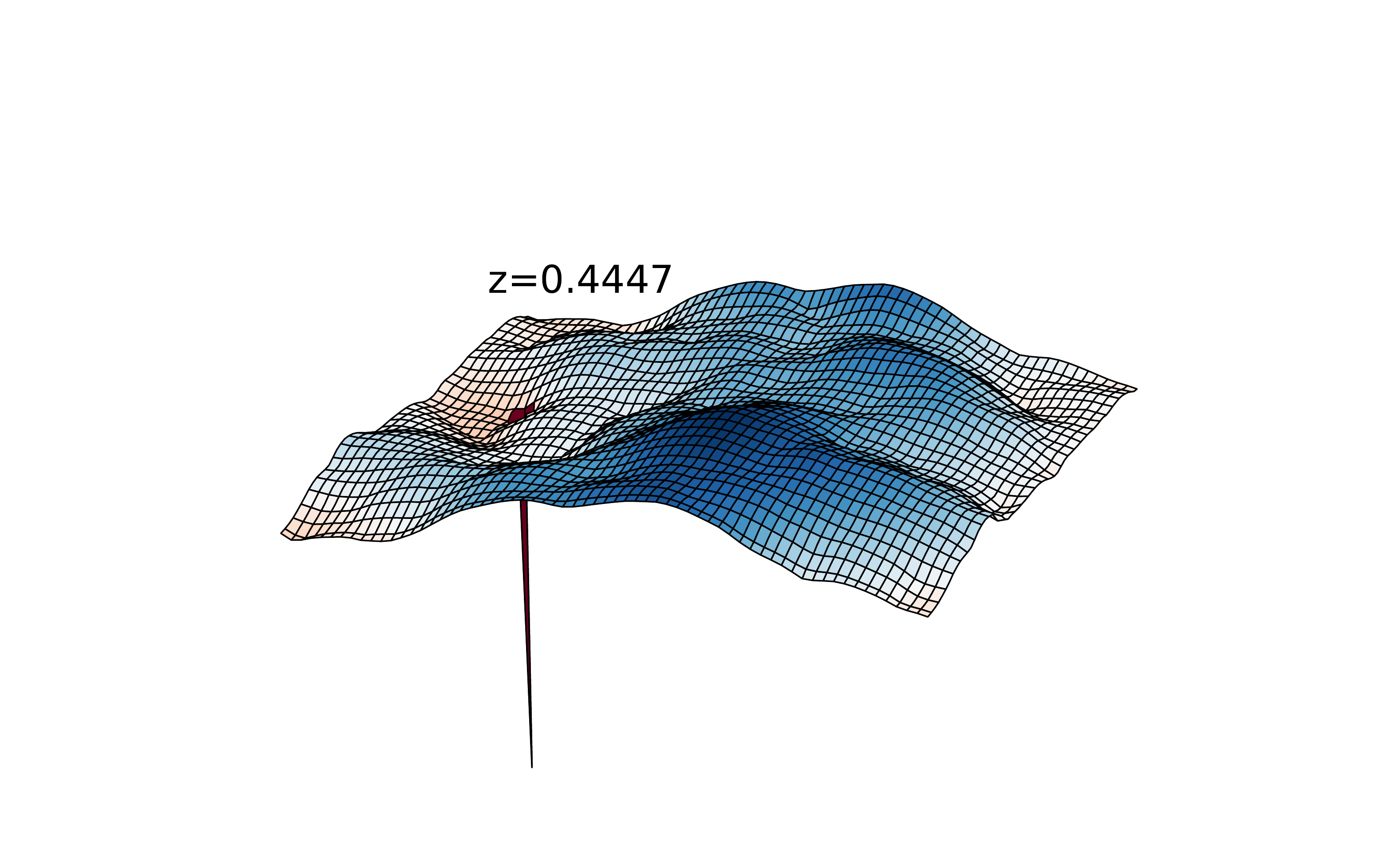} }}\hfill
   {{\includegraphics[scale=0.15, trim=-10cm 2cm 0cm -1cm]{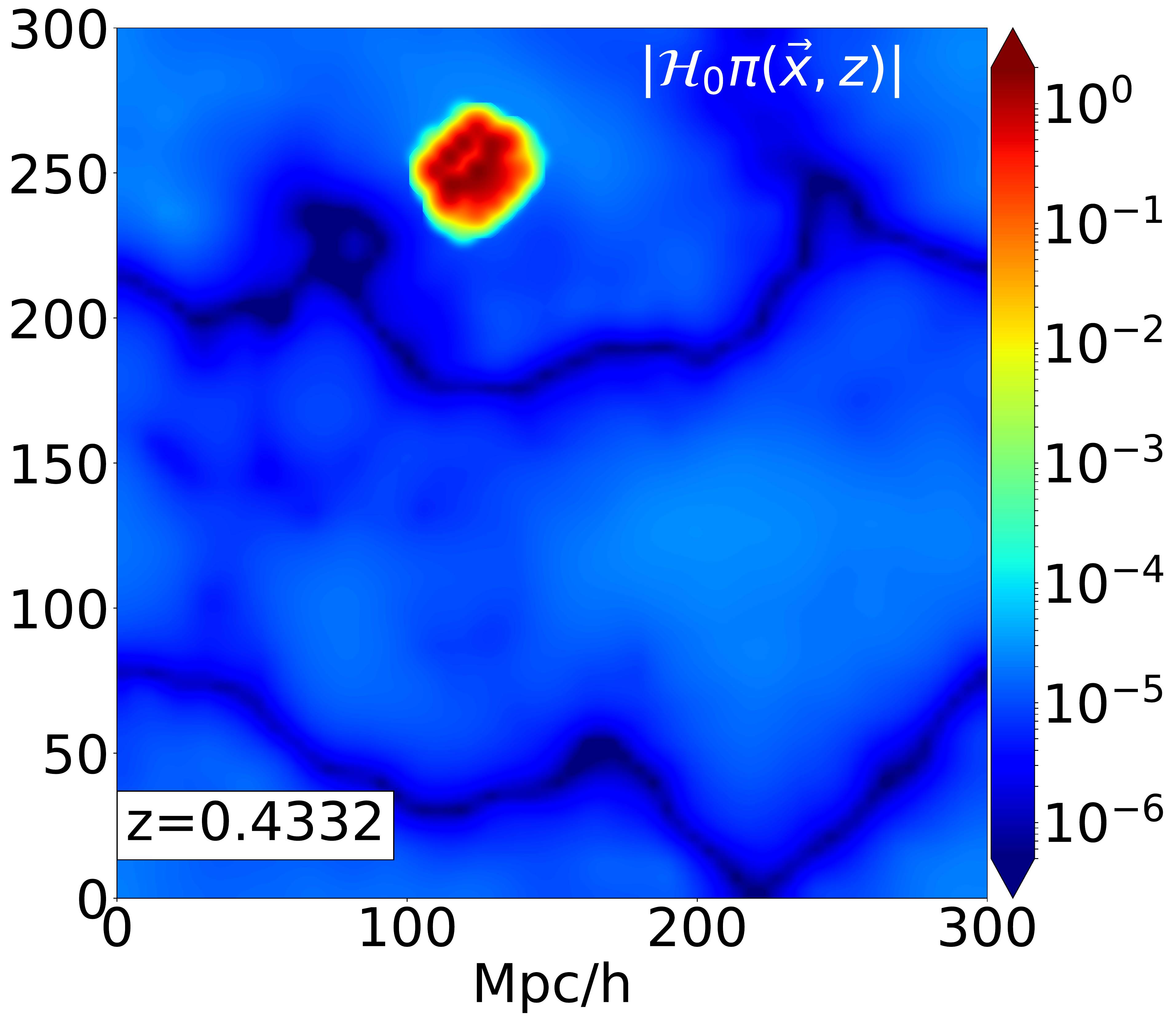} }}
      %%%%%%%%%%%%%%%%%%%%%%
\caption{The evolution of the scalar field $\pi$ in time, increasing from top to bottom (note that left and right correspond to different time scales). \emph{Left:} 3D plot of the scalar field in a $x-z$ cross-section of the simulation, where its $y$ value corresponds to the point with highest curvature of the potential. \emph{Right:} A color map of the absolute value of the scalar field $|\mathcal{H}_0 \pi|$ at different redshifts, for the same cross-section. Around the blow-up time the instability is formed locally in the point with the maximum curvature which physically corresponds to the center of the dark matter halo with highest density.
}
\label{1D_snapshot}
 \end{minipage}
 \end{figure*}

Starting the simulation at some initial time and solving the full equations of motion in the weak-field approximation for low speed of sound, one finds numerically that the scalar field $\pi$ diverges and as a result the simulation breaks down in finite time. In Fig.~\ref{1D_snapshot} on the right we show  the absolute value of the scalar field 
in the $x-z$ plane taken at the $y$-position of the point with the maximal second derivative of $\pi$. On the left side we render the same 2D section as a 3D plot, where the height shows the value of the field for better illustration. In these images we see how, over a short period of time in the simulation, an instability is formed around the minimum with largest second derivative and blows up. Since other 
quantities are coupled to the scalar field, like for example the gravitational potential, they 
will also diverge 
and eventually the simulation breaks down.
The instability is local and thus if we look at 
regions far away from
the point with maximal second derivative of the scalar field, at the same redshift we see no hint of instability until the simulation itself fails. The reason why the instability is first formed around the minimum with highest curvature will become clear when we study the system in %1+1 D
a simplified symmetric setup
analytically in Sec.~\ref{sec:one_d}.
The worked example
shown in Fig.~\ref{1D_snapshot} is for illustrative purposes and is obtained %using cosmological $N$-body
{from a}
simulation with %$h=0.67556$, $\omega_b=0.022032$, $\omega_\text{cdm}=0.12038$, $T_\text{CMB}=2.7255$,
$c_s^2 = 10^{-7}$ and $w = -0.9$.
%%%%%%%%%%%%%

\begin{figure}[tb]%
    \centering
    \hspace*{-1cm}  
   {{\includegraphics[scale=0.4]{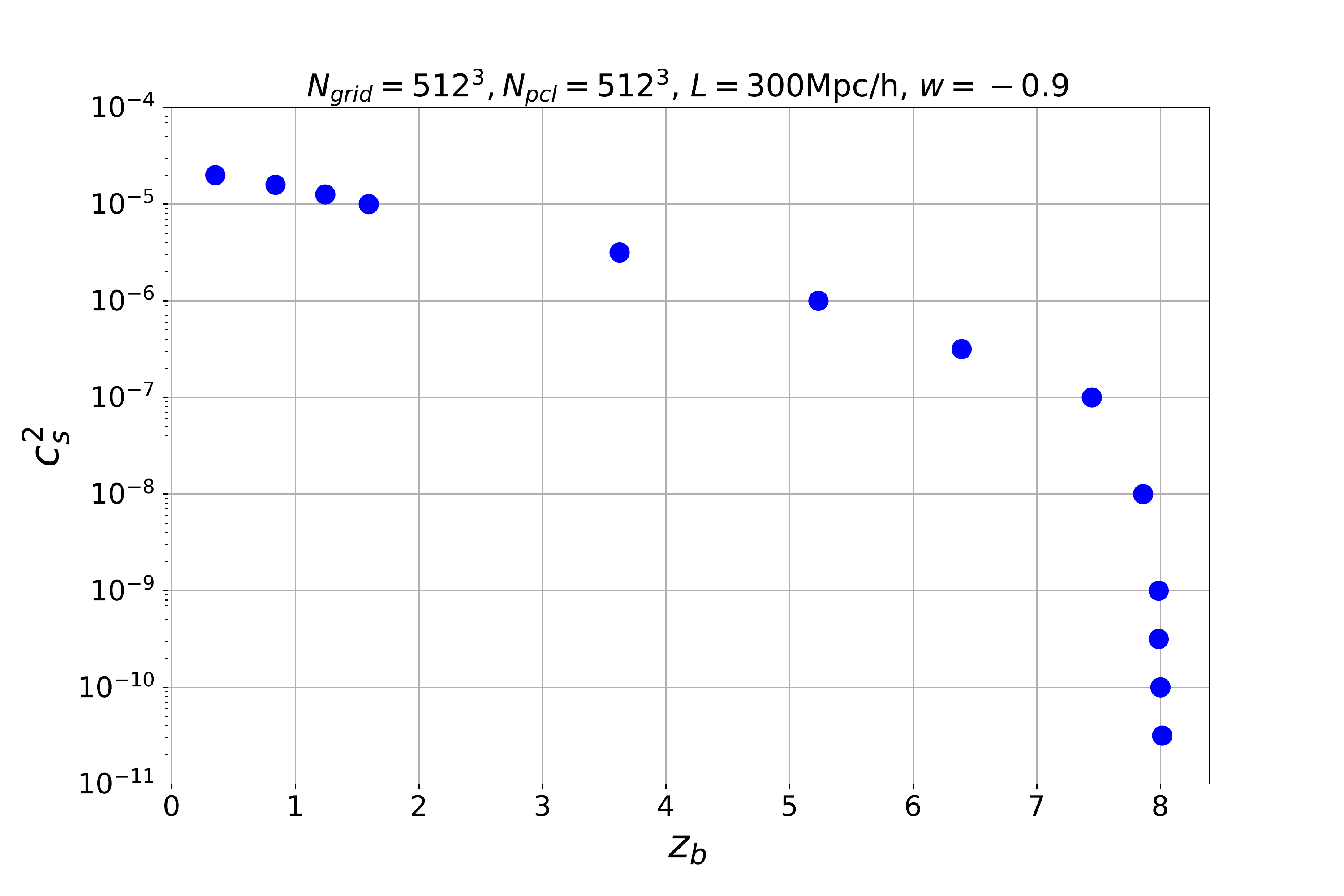} }}%
        \caption{ The speed of sound squared as a function of blow-up redshift. There are two limits, namely high speed of sound for which the system does not blow-up, and very low speed of sound when the system blows up at a redshift close to the blow-up redshift for $c_s^2 \to 0$. It is worth mentioning that the blow up redshift depends on the resolution of simulation (see Fig.~\ref{spatial_resolution_v2}), which also affects somewhat the minimal speed of sound for which the system is stable.} 
    \label{speed_vs_zb}%
\end{figure}

Studying the behaviour for different values of the speed of sound systematically,
using \kev\ with the full implementation of clustering DE according to Eqs. \eqref{zeta_eq1} and \eqref{zeta_eq2}, we find that the solutions are unstable and diverge in finite time for low speed of sound only. Indeed, for otherwise fixed cosmological parameters and $w_0 =-0.9$, our numerical studies indicate that the system only blows up when $c_s^2 \lesssim 10^{-4.7}$. 
In  Fig.~\ref{speed_vs_zb} we show the ``blow-up redshift'' $z_b$ 
for different speeds of sound for a fixed resolution of $0.58$ Mpc/h. As the figure suggests, when increasing $c_s^2$ there is a critical value for $c_s^2$ where the system becomes stable.

In Appendix~\ref{convergence_tests} we discuss the effect of precision parameters (temporal and spatial resolution) on our results. We show in Appendix \ref{sec:timeprec} that increasing the time resolution of the solver does not change the blow-up time significantly. 
The dependence of our results on the spatial resolution is interesting and can be traced back to the fact that increasing the resolution also enhances the maximum amplitude of perturbations in the initial conditions, discussed further below.
This dependence can be understood quantitatively using the extreme value theorem, as we discuss in Appendix~\ref{app_spatial}.

We study the robustness of the critical $c_s^2$ on the chosen cosmology in Appendix~\ref{cs_app} where we show results from simulations that remain matter dominated forever. We find that the limit for the stability of the equation is not changed significantly even if we let the simulations run far into the future. This rules out the possibility that high speed of sound simulations are only stable due to the limited time available until the gravitational potential starts to decay when matter domination ends and DE takes over.

\begin{figure}[tb]%
    \centering
    \hspace*{-1cm}  
   {{\includegraphics[scale=0.4]{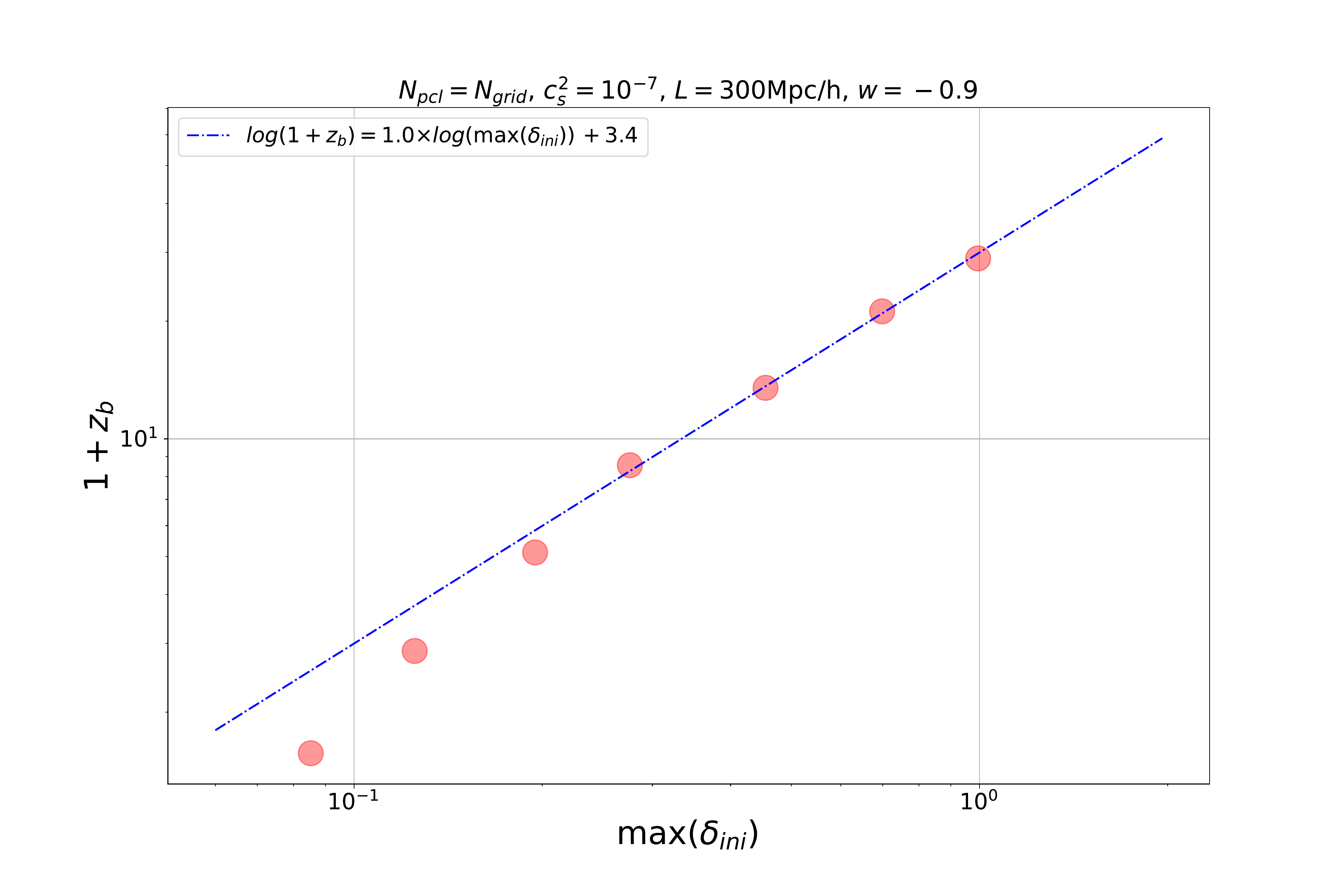} }}%
        \caption{ The blow-up redshift as a function of maximum initial density. The data correspond to cosmological $N$-body simulations for $N_\mathrm{grid} = N_\mathrm{pcls} = 64^3$, $128^3$, $256^3$, $512^3$, $1024^3$, $2048^3$, $3600^3$, where we compute  $\max(\delta_\mathrm{ini})$ on the lattice for each simulation. The dashed line represents the expected scaling in matter domination which shows a linear relation between $1+z_b$ and $\max(\delta_\mathrm{ini})$.}
    \label{density_z_b}%
\end{figure}

In Fig.~\ref{density_z_b} we present the dependence of blow-up redshift on the initial conditions of the simulation. 
As we are going to explain in detail in Appendix~\ref{app_spatial}, for a fixed box size increasing the resolution of the grid and number of particles would result in increasing the initial density of the perturbations as we are probing smaller scales. But since the blow-up happens first at the point with the highest curvature of the potential, corresponding to the point with the highest density, this in turn affects the blow-up redshift.
Our results show that  in matter domination there is a linear relation between the blow-up redshift and the initial density. In the next section we will validate this proportionality using a simplified spherically symmetric setup. 
In Appendix~\ref{app_spatial} we discuss the relation between $\langle \max (\delta) \rangle$ and the resolution of a  simulation.

Interestingly we also know that the limit of low speed of sound is the limit where important linear terms in the dynamics of {the} scalar field {are suppressed} and we end up with a highly non-linear evolution of the field. The critical value $c_s^2 \sim 10^{-4.7}$ in Fig.~\ref{speed_vs_zb} can be understood by comparing the two most important terms in the dynamics of the scalar field. As we are going to show, stability of this system is ensured when the term $c_s^2 \nabla^2 \pi $ dominates over the non-linear term $-^1\!/_{\!2}\,\HH \big(5c_s^2  + 3w  -2\big) \; {(\vec\nabla \pi)^2}$. {Following the spirit of the weak-field expansion mentioned earlier, we can use a dimensional analysis and write} 
$\nabla^2 \pi \sim L^{-2} \pi$ and $(\vec\nabla \pi)^2 \sim L^{-2} \pi^2${, where $L$ is a characteristic length scale}.

Furthermore, in matter domination we have %i.e., 
$\HH = 2/\tau$, and the first-order perturbative solution for $\pi \approx {}^1\!/_{\!3}\,\Psi \tau$ calculated in Eq.~(D.6) in \cite{Hassani:2019lmy}. {The two terms will therefore be of similar size if the speed of sound fulfils the relation} 
\be
c_s^2 \approx  
\big( \frac23-w -{\frac 5 3 c_s^2}\big) \Psi\,.
\ee
With $w < 0$ this relation is satisfied when $c_s^2 \sim \Psi$, and in cosmology we typically have $\Psi \sim 10^{-5}$ which gives a critical value of $c_s^2$ commensurate with our numerical measurement.

Given the fact that the two relevant terms of the equations of motion are independent of the gravitational coupling, the instability is not a result of scalar field self-gravity. This can be checked numerically by turning off the scalar field's contribution to the gravitational field equations (the stars in Figs.~\ref{temporal_resolution} and \ref{spatial_resolution_v2} as well as the discussion in Appendix~\ref{app_spatial}), effectively turning it into a spectator field. Even in this case we find that the perturbations in the scalar field, sourced by the gravitational potentials of dark matter alone, become unstable at almost the same time as in the case with full gravitational coupling. 
Motivated by these numerical results 
in the cosmological context we study a simplified version of the full equations analytically in a symmetric setup.

%%%%%%%%%%%%%%%%%
%%%Section  : Analytical results
%%%%%%%%%%%%%%%%%
\section{Analytical results \label{sec:one_d}}
In this section we discuss the most important terms in the PDE governing the scalar field dynamics. In particular, we discuss  the non-linear dynamics in 1+1 dimensions if we consider either spherically symmetric or plane-symmetric solutions. We show that the non-linear PDE is suffering from an instability with similar behavior of what we found in the realistic 3+1 dimensional case. Studying the full dynamics even in 1+1 D remains a difficult task and  
is beyond the scope of this paper. However, the specific term in the equation of motion which we have recognized as the root cause of the instability of the system, namely $(\vec\nabla \pi)^2$ \cite{Hassani:2021tdd}, is studied thoroughly in \cite{PanShi_scale_invariant,PanShi_stability,PanShi_Hamilton, JPE_Hassani} in a mathematical context. Here, we study this term with a more physical  approach.

We rewrite Eq.~\eqref{zeta_eq2} as a second-order PDE which is more appropriate for analytical studies, 
\begin{multline}
\partial_{\tau}^2\pi  + \HH (1-3w) \partial_{\tau}\pi  + \Big( \partial_{\tau} \HH - 3w \HH^2 + 3 c_{s}^{2} (\mathcal{H}^{2} - \partial_{\tau} \HH)  \Big)  \pi   \\ 
-  \partial_{\tau}\Psi +3\HH (w- c_s^2) \Psi - 3 c_{s}^{2} \partial_{\tau}\Phi-c_{s}^{2}  \nabla^2\pi  = \mathcal{N}(\pi,  	\partial_\tau \pi,  \vec\nabla \pi,  \partial_{\tau} \vec\nabla \pi, \nabla_i \nabla_j \pi)  \,, \label{full_1Deq}
\end{multline}
where $  \mathcal{N}(\pi,  	\partial_\tau \pi,  \vec\nabla \pi,  \partial_{\tau} \vec\nabla \pi, \nabla_i \nabla_j \pi)$ includes all the non-linear terms,
\begin{multline}
\mathcal{N}(\pi,  	\partial_\tau \pi,  \vec\nabla \pi,  \partial_{\tau} \vec\nabla \pi, \nabla_i \nabla_j \pi) =    - \frac{\HH}{2}  \big(5c_s^2  + 3w  -2\big) \; {(\vec\nabla \pi)^2}
 	+ 2(1- c_s^2)  \vec\nabla \pi \cdot \vec\nabla\partial_{\tau}  \pi
	 \\
	 - \Big[ (c_s^2-1) \big( \partial_{\tau}  \pi + \HH \pi -\Psi \big)+ c_s^2 (\Phi - \Psi) + 3 \HH c_s^2 (1+w)\pi  \Big] \nabla^2 \pi 
	 + (2 c_s^2 -1)  \vec\nabla \Psi \cdot \vec\nabla \pi   	  
	 \\
	  - c_s^2 \vec\nabla \Phi \cdot \vec\nabla \pi 
 +\frac{c_s^2 -1 }{2} \vec\nabla \cdot \left( (\vec \nabla \pi)^2 \vec\nabla \pi\right)  \,.
\end{multline} 
This equation is called a ``non-linear damped wave equation" \cite{GALLAY199842} in the mathematics literature. It has been studied by mathematicians in 1+1 D for some types of non-linearities, mainly of the form $ \mathcal{N}(\pi, \partial_\tau \pi) $ \citep{hayashi2004,1998JDE...150...42G,  2016arXiv160504616I} but not in the general form appearing in the EFT of DE which is  $\mathcal{N}(\pi,\partial_\tau \pi,\vec \nabla \pi, \nabla_i \nabla_j \pi)$. In fact, the important remark is that for large speeds of sound and using the fact that $\pi$, $\partial_\tau \pi$ and $\vec \nabla \pi$ are small, the dominant term in the dynamics would be the linear part of the PDE, whereas in the limit $c_s^2 \to 0$ the term $c_s^2 \nabla^2 \pi$, which is the restoring force in the linear wave equation, vanishes and therefore the non-linear terms become relevant.

In this limit, \ie\ $c_s^2 \to 0$, the equation also has a new symmetry whenever the scale factor is a power law in $\tau$ and therefore $\mathcal{H} \propto \tau^{-1}$: it becomes invariant under the rescaling $\Psi \to \lambda^2 \Psi$, $\pi \to \lambda \pi$, $\tau \to \lambda^{-1} \tau$. The scale invariance also approximately holds at finite $c_s^2$ using $c_s \to \lambda c_s$ as long as $c_s^2$ remains sufficiently small under the rescaling, \ie\ for scales much larger than the sound horizon. Scale invariance is indeed expected if the physical problem lacks a characteristic scale such as a sound horizon or a break in the power law in $\tau$. This immediately leads to an interesting conclusion if we consider $\pi$ as a spectator field in matter domination where it would be sourced by the gravitational potential $\Psi$ produced by non-relativistic matter. If we assume that $\Psi$ is independent of time (a good approximation when matter is in the linear regime) and have a solution $\pi(\tau; \Psi)$, we can generate new solutions $\pi(\tau; \lambda^2 \Psi) = \lambda^{-1} \pi(\lambda \tau; \Psi)$. Evidently, if for a given gravitational source field $\Psi$ the spectator field $\pi$ diverges at a certain value $z_b + 1 \propto \tau_b^{-2}$, that value is directly proportional to the overall amplitude of $\Psi$, \ie\ $z_b + 1 \propto \Psi$. Here we use the fact that $z + 1 \propto \tau^{-2}$ in matter domination. As we will see shortly, the divergence is actually sensitive to the maximum curvature of $\Psi$ which is of course proportional to $\Psi$ itself in these scaling solutions with constant $\lambda$.

%%%%%%%%%%%%%%%%%
%%%SUBSection  : Spherically symmetric case
%%%%%%%%%%%%%%%%%
\subsection{Spherical symmetry \label{Spherically}}
In this subsection we study the PDE \eqref{full_1Deq} in the %limit $c_s^2 \to 0$ 
regime $c_s^2 \ll 1$
assuming spherical symmetry.  
We start from Eq.~\eqref{full_1Deq} and assume a spherically symmetric scenario, \ie\ all fields are functions of $\tau$ and $r$ only. Furthermore, we choose $N = \ln a$ as the new time coordinate, and also rescale $\pi = \tilde{\pi} / \mathcal{H}$. The equation reads  
\begin{multline}
\label{eq:sphericalPDE}
\partial^2_N \tilde{\pi} + \left(1 - 3 w - \partial_N \ln\mathcal{H}\right) \partial_N \tilde{\pi} + \left[3 w \left(\partial_N \ln\mathcal{H} - 1\right) - \partial^2_N \ln\mathcal{H}\right] \tilde{\pi} {- \frac{c_s^2}{\mathcal{H}^2} \left(\frac{2}{r} \partial_r \tilde{\pi} + \partial_r^2 \tilde{\pi}\right)} \\
 + 3 w \Psi - \partial_N \Psi = \frac{2 {-} 3 w {-4 \partial_N \ln \mathcal{H}}}{2 \mathcal{H}^2} \left(\partial_r \tilde{\pi}\right)^2 {- \frac{\partial_r \Psi - 2 \partial_r \partial_N \tilde{\pi}}{\mathcal{H}^2}\partial_r \tilde{\pi}} \\{+ \frac{\left(1 - \partial_N \ln\mathcal{H}\right) \tilde{\pi} + \partial_N \tilde{\pi} - \Psi}{\mathcal{H}^2 r} \left(2 \partial_r \tilde{\pi} + r \partial_r^2\tilde{\pi}\right) - \frac{\left(\partial_r \tilde{\pi}\right)^2 \left({2} \partial_r \tilde{\pi} + {3} r \partial^2_r \tilde{\pi}\right)}{{2} \mathcal{H}^4 r}}\,,
\end{multline}
where we neglect $c_s^2$ against coefficients of order unity (including $w$) and assume $\Psi \sim \Phi$ in that context. In other words, the only term for which the value of $c_s^2$ is important (once it is assumed that $c_s^2 \ll 1$) is the linear restoring force.
Note that if we neglect radiation in the late Universe, the Hubble function is
\be 
 \mathcal{H}^2 = H_0^2 \left[\Omega_m e^{-N} + \left(1-\Omega_m\right) e^{-(1 + 3 w)N}\right]\,,
\ee 
such that
\be
\partial_N \ln\mathcal{H} = \frac{\partial_N \left(\mathcal{H}^2\right)}{2 \mathcal{H}^2} = -\frac{\Omega_m + \left(1 + 3 w\right) \left(1 - \Omega_m\right) e^{-3 w N}}{2 \Omega_m + 2 \left(1 - \Omega_m\right) e^{-3 w N}}\,.
\ee
Moreover, if we neglect gravitational backreaction from the scalar field $\pi$ and assume that $\Psi$ is generated by a linear matter perturbation, we can also write the time evolution equation for $\Psi$,
\be 
 \partial_N^2 \Psi + \left(3 + \partial_N \ln \mathcal{H}\right) \partial_N \Psi + \left(2 - \frac{3\Omega_m}{2\Omega_m + 2\left(1 -  \Omega_m\right)e^{-3 w N}} + \partial_N \ln\mathcal{H}\right) \Psi = 0\,. \label{psi_eq}
\ee 
This equation can be solved analytically in terms of hypergeometric functions. However, in order to gain more analytic insight it is instructive to consider the case of matter domination, \ie\ $\Omega_m = 1$. In this limit we find $\mathcal{H}^2 = H_0^2 e^{-N}$, $\partial_N \ln \mathcal{H} = -^1\!/_{\!2}$, and the linear gravitational potential $\Psi$ is constant in time. Eq.~\eqref{eq:sphericalPDE} simplifies to
\begin{multline}
\label{eq:sphericalPDEmatter}
    \partial^2_N \tilde{\pi} + \frac{3}{2}\left(1 - 2 w\right) \partial_N \tilde{\pi} - \frac{9}{2} w \tilde{\pi} {- \frac{c_s^2}{H_0^2} \left(\frac{2}{r} \partial_r \tilde{\pi} + \partial_r^2 \tilde{\pi}\right) e^N} + 3 w \Psi = \frac{4 - 3 w}{2 H_0^2} \left(\partial_r \tilde{\pi}\right)^2 e^N  \\
     - \frac{\partial_r \Psi - 2 \partial_r \partial_N \tilde{\pi}}{H_0^2} \partial_r \tilde{\pi} e^N + \frac{3 \tilde{\pi} + 2\partial_N \tilde{\pi} - 2\Psi}{2H_0^2 r} \left(2 \partial_r \tilde{\pi} + r \partial_r^2 \tilde{\pi}\right) e^N - \frac{\left(\partial_r \tilde{\pi}\right)^2 \left({2} \partial_r \tilde{\pi} + {3} r \partial^2_r \tilde{\pi}\right)}{{2} H_0^4 r}e^{2N}\,.
\end{multline}
We can easily infer that the nonlinear right-hand side is exponentially suppressed at early times as $N \to -\infty$ and therefore the initial solution should approach its linear expression $\tilde{\pi} \to {}^2\!/_{\!3}\,\Psi$ for scales larger than the sound horizon. Note in particular that the linear solution (for $w < 0$) would be stable at all times if the non-linear self-coupling of $\tilde{\pi}$ is neglected, as was the case for previous numerical studies mentioned in Sec.~\ref{sec:intro}.

Further insight can be found if we consider the solution close to an extremum of the potential. Let us write
\begin{equation}
    \Psi(r) = \Psi_0 + \frac{1}{2} r^2 H_0^2 \Psi_2\,,
\end{equation}
where the factor $H_0^2$ is introduced to render the coefficient $\Psi_2$ dimensionless. We can write the solution for $\tilde{\pi}$ as the asymptotic one plus a correction, \ie
\begin{equation}
    \tilde{\pi}(N, r) = \frac{2}{3} \Psi_0 + \frac{1}{3} r^2 H_0^2 \Psi_2 {+ \frac{4 c_s^2}{5 - 15 w} \Psi_2 e^N} + \epsilon_0(N) + \frac{1}{2} r^2 H_0^2 \epsilon_2(N)\,. \label{eq:pitildeansatz}
\end{equation}
Inserting these ans\"atze into Eq.~\eqref{eq:sphericalPDEmatter} we find that it neatly separates into two independent parts, one that has no $r$-dependence,
\begin{equation}
{\partial_N^2 \epsilon_0 + \frac{3}{2} \left(1 - 2 w\right) \partial_N \epsilon_0 - \frac{9}{2} w \epsilon_0 = \frac{1}{2} \left(2 \Psi_2 + 3 \epsilon_2\right) \!\left(\!3 \epsilon_0 + 2 \partial_N \epsilon_0 + \frac{4 c_s^2}{1- 3 w} \Psi_2 e^N\!\right) e^N + 3 c_s^2 \epsilon_2 e^N,}\label{eq:epsilon0}
\end{equation}
and one that scales as $r^2$ and reads
\begin{multline}
\label{eq:epsilon2}
    \partial_N^2 \epsilon_2 + \frac{3}{2} \left(1 - 2 w\right) \partial_N \epsilon_2 - \frac{9}{2} w \epsilon_2 = \left(\frac{2}{3} \Psi_2 + \epsilon_2\right)\! \left[\frac{1}{3}\left(2 - 6 w\right) \Psi_2 + \frac{1}{2}\left(17 - 6 w\right) \epsilon_2 + 7 \partial_N \epsilon_2\right] e^N\\
    - {5} \left(\frac{2}{3} \Psi_2 + \epsilon_2\right)^3 e^{2 N}\,.
\end{multline}
The second equation is independent of the value of $c_s^2$. The asymptotic solution for $\epsilon_2$ at $N \to -\infty$ can be inferred by recognising that $\epsilon_2$ is of higher perturbative order than $\Psi_2$ and therefore becomes subdominant in the nonlinear contribution. One finds that $\epsilon_2 \to {}^8\!/_{\!45}\,e^N \Psi_2^2$ as $N \to -\infty$. At this point it is also worth noting that this asymptotic solution is entirely governed by the first two non-linear terms on the right-hand side of Eq.~\eqref{eq:sphericalPDEmatter}, \ie\ the terms quadratic in gradients. The other two terms are asymptotically subdominant, and hence do not efficiently trigger the non-linear evolution of $\tilde{\pi}$. This justifies our claim that the term $(\vec\nabla \pi)^2$ is the most relevant non-linear term when discussing the instability. The other terms do, however, have a small effect on the precise time when the divergence of $\epsilon_2$ occurs.

\begin{figure}[tb]
    \centering
    \hspace*{-1cm}  
   {{\includegraphics[width=0.8\textwidth]{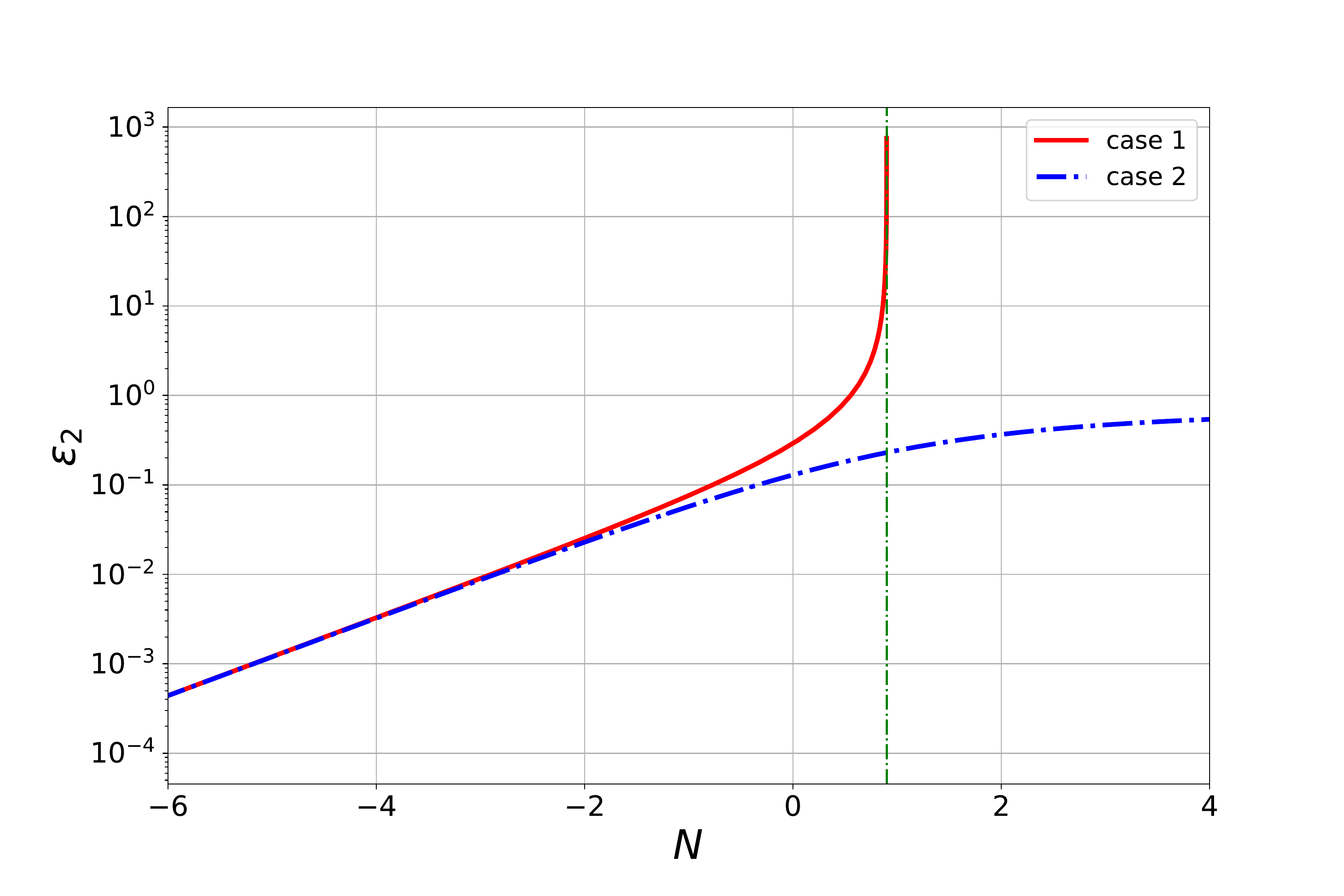} }}%
        \caption{Fully non-linear solution of $\epsilon_2$ for $\Psi_2 = 1$ (case 1, red) and $\Psi_2 = -1$ (case 2, blue). The divergence occurs only in case 1, at approximately $N = 0.9$ (dashed green line). The equation of state is chosen as $w=-0.9$.}
    \label{fig:epsilon2}
\end{figure}

In Fig.~\ref{fig:epsilon2} we show the fully non-linear solution for $\epsilon_2$ in the two cases where $\Psi_2 = \pm 1$. A divergence only occurs in the case where $\Psi_2$ is positive (around minima of the potential wells). For $\Psi_2 < 0$ the solution as $N \to +\infty$ approaches the exact solution $\epsilon_2 = -^2\!/_{\!3}\,\Psi_2 -\sqrt{-\Psi_2}\,e^{-N/2} + {}^1\!/_{\!2}\,e^{-N}$ such that the curvature of $\tilde{\pi}$ vanishes asymptotically. However, it decays exactly as fast as $\mathcal{H}$ so that the curvature of $\pi = \tilde{\pi} / \mathcal{H}$ becomes constant.

The scale invariance manifests itself in the fact that Eq.~\eqref{eq:epsilon2} is invariant under the transformation $\Psi_2 \to \lambda^2 \Psi_2$, $\epsilon_2 \to \lambda^2 \epsilon_2$, $N \to N - 2 \ln \lambda$. Since $\Psi_0$ does not appear in the nonlinear equations for $\epsilon_0$, $\epsilon_2$, it is clear that the nonlinear instability is indeed governed by the curvature of the potential $\Psi$, \ie\ the amplitude of the coefficient $\Psi_2$ which is also directly proportional to the matter density contrast. The scaling symmetry implies that the redshift at which the divergence occurs is proportional to this coefficient.

The linear relation between the blow-up redshift and the maximal initial density in matter domination agrees with the results from the three-dimensional simulations of Sec.\ \ref{sec_3_numerics}, as shown in Fig.\ \ref{density_z_b}, indicating that the analytically tractable spherically symmetric case is able to capture the most important features of the divergence dynamics.

While it might appear that the blow-up occurs independent of the value of $c_s^2$ since the solution of $\epsilon_2$ does not depend on it, in a more realistic setting the evolution of the linear solution does matter. The correction due to $c_s^2$ acts to increase/decrease the value at the minimum/maximum of $\tilde{\pi}$, which under more realistic boundary conditions tends to decrease all derivatives.
As a crude estimate we could say that this effect becomes important when the time-dependent term of the linear solution becomes $\sim |\Psi_0|$, which occurs roughly at $N \sim -2 \ln c_s - \ln \Psi_2 + \ln |\Psi_0| + 1.6$.
The numerical constant is not very important and was computed assuming $w \simeq -1$. We may then argue that the critical speed of sound is given by the estimate $N_b \sim -2 \ln c_s - \ln \Psi_2 + \ln |\Psi_0| + 1.6$, where $N_b$ is the value of $N$ at which $\epsilon_2$ blows up.
From Fig.~\ref{fig:epsilon2} and using the approximate scale invariance we infer that $N_b + \ln \Psi_2 \sim 0.9$ independent of the value of $\Psi_2$. This results in an estimate for the critical speed of sound that is very much in agreement with our previous estimate given in Sec.~\ref{sec_3_numerics}. This argument of course only holds as long as the initial assumption that $c_s^2 \ll 1$ is valid. 

With the asymptotic solution of $\epsilon_2$ the corresponding asymptotic solution of $\epsilon_0$ from Eq.~\eqref{eq:epsilon0} as $N \to -\infty$ reads
\begin{equation}
    \epsilon_0 \to \frac{8}{105} c_s^2 \frac{17 - 6 w}{2 - 9 w (1 - w)} e^{2 N} \Psi_2^2\,.
\end{equation}
This solution
reaches a value similar to the time-dependent term of the linear solution when $N + \ln \Psi_2 \sim 0.8$, which always happens close to $N = N_b$ in the case where a blow-up occurs. This means that considering $\epsilon_0$ will not change our estimate of the critical speed of sound in a significant way. On the other hand, due to its coupling to $\epsilon_2$, $\epsilon_0$ grows without bounds towards $N_b$. This raises the question whether under more realistic boundary conditions, following the thinking of the previous paragraph, the instability could actually be halted.

Related to this question it is interesting to note that the PDE~\eqref{full_1Deq} has a particular exact solution for $c_s^2 > 0$ if we drop the last two terms on the right-hand side (which are often very subdominant). This can be seen by making the ansatz $\tilde{\pi} = C e^N + {}^2\!/_{\!3}\,\Psi$, for which the corresponding equation becomes
\begin{equation}
    \frac{15 H_0^2}{4} \left(1 - 3 w\right) C - c_s^2 \left(\frac{2}{r}\partial_r \Psi + \partial^2_r\Psi\right) = \left(\frac{1}{3} - w\right) \left(\partial_r \Psi\right)^2\,.
\end{equation}
A valid solution for $\Psi$ can be obtained with the Hopf-Cole transformation $\Psi = 3 c_s^2 \ln v / (1-3 w)$, for which we get
\begin{equation}
    \frac{5 H_0^2}{4 c_s^4} \left(1-3w\right)^2 C v - \frac{2}{r} \partial_r v - \partial^2_r v = 0\,.
\end{equation}

A regular solution for which the Hopf-Cole transformation remains defined globally is $v(r) = D \sinh(k r) / (k r)$,  
with $k^2 = {}^5\!/_{\!4} H_0^2 (1-3w)^2 C / c_s^4$ and $C>0, D>0$. While this is of course a highly fine-tuned solution, it is interesting to note that the radial profile of $\tilde{\pi}$ does not evolve in this potential. Since we can fix $C$ and $D$ in a way to give any desired second-order Taylor expansion around the minimum of $\Psi$, this shows that for any $c_s^2 > 0$ there exists a local solution where the gradients of $\tilde{\pi}$ do not evolve and therefore do not lead to a blow-up. Under general initial conditions it remains an open question whether the singularity can be avoided in this way.

%%%%%%%%%%%%%%%%%
%%%SUBSection  : Planar symmetric
%%%%%%%%%%%%%%%%%

\subsection{Planar symmetry \label{Planar}}
In this subsection we study the blow-up dynamics in more detail, using an even more simplified
version of Eq.~\eqref{full_1Deq} assuming planar symmetry in a non-cosmological setup,
\begin{align}
 \partial_{\tau}^2\pi(\tau,x)  = c_s^2 \nabla^2 \pi(\tau,x) + \alpha \bigl(\nabla \pi(\tau,x)\bigr)^2 \, .\label{eq_oneD}
\end{align}
This equation is written following our previous study \cite{Hassani:2021tdd} and the mathematical studies \citep{PanShi_scale_invariant,PanShi_stability,PanShi_Hamilton, JPE_Hassani} where in addition to the non-linear term we consider the  linear term $c_s^2 \nabla^2 \pi$. We therefore consider the two important terms in the dynamics of the scalar field, \ie\  the instability part $(\nabla \pi)^2$ and the pressure term which stabilises the system. 

\begin{figure*}
\begin{minipage}{\textwidth}
\includegraphics[width=1.0\textwidth]{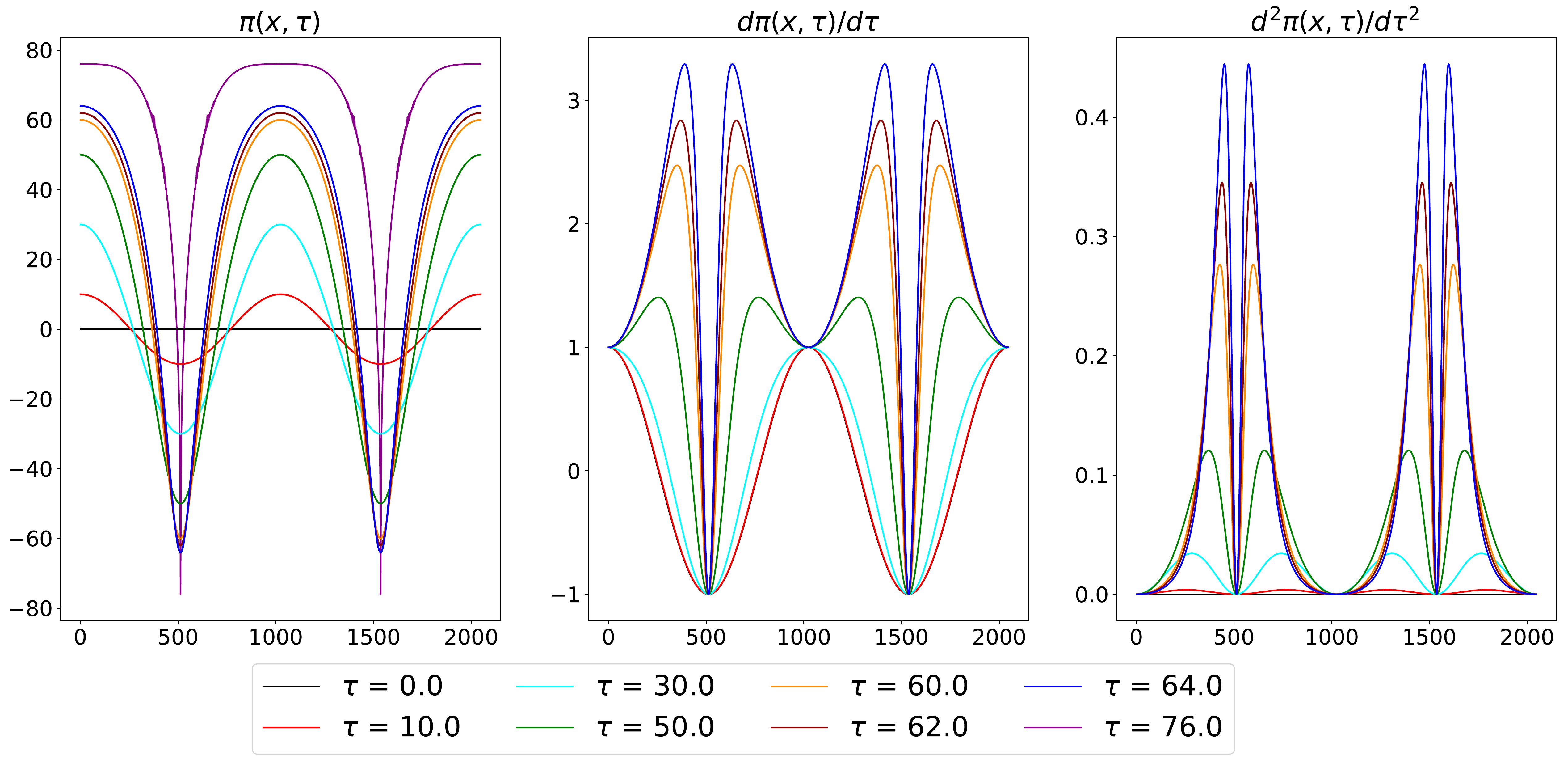}\hfill
\includegraphics[width=1.0\textwidth]{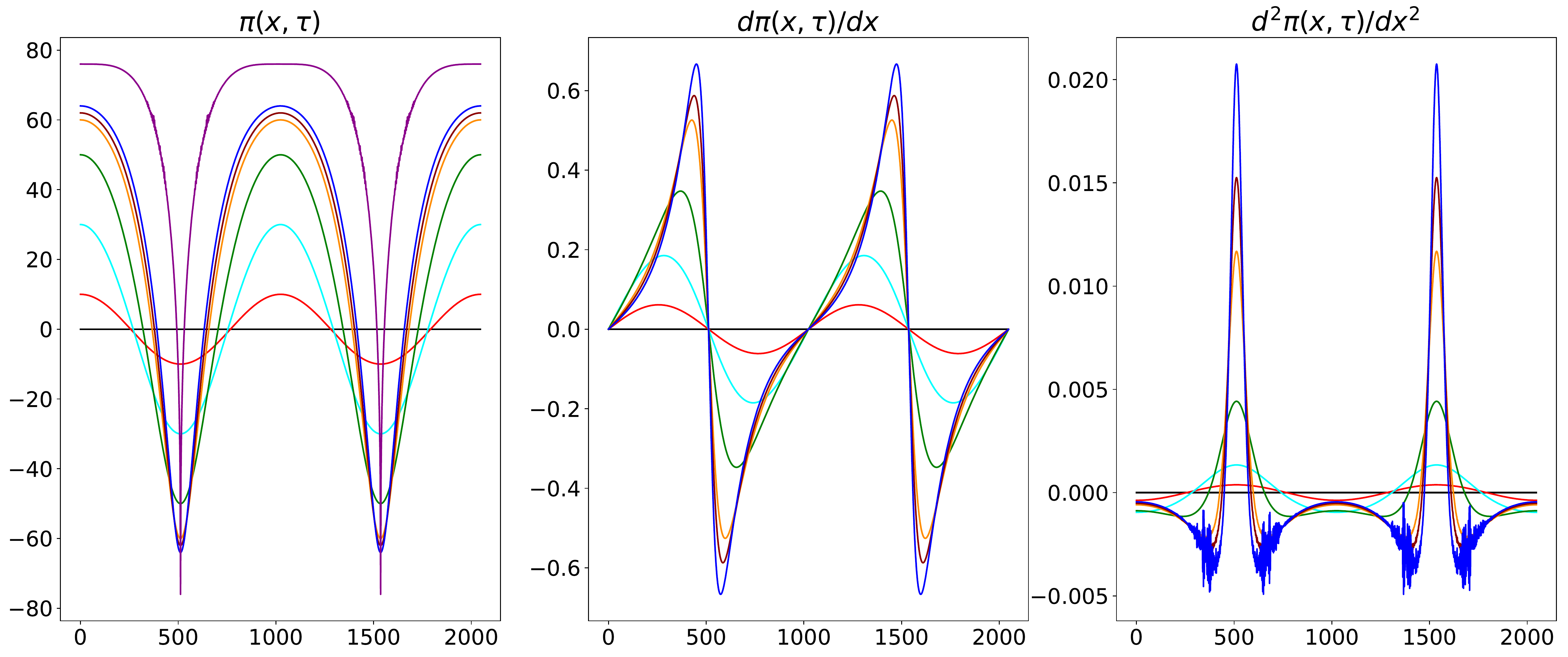}\hfill
\caption{  {\emph{Top}}: The scalar field and its time derivatives on the 1+1 D lattice at different times  are shown, using the potential $\Psi = \cos(4 \pi x / L)$ as an initial condition. Analytically the curvature of the scalar field at the minimum blows up at time $\tau_b \approx 76$.
\newline
 {\emph{Bottom:}} The same scalar field and its spatial derivative on the lattice for different times are shown. Due to the numerical noises appearing in the derivatives of the field we only show the results for the derivatives up to $\tau =64$. It is also interesting to see that $\partial_x \pi$ behaves similar to a gradient catastrophe that one would see in some situations in fluid dynamics.}
\label{Numerics_1D_blowup}
 \end{minipage}
 \end{figure*}

In the limit $c_s^2 \to 0$ and 
rescaling $\pi \to \pi/\alpha$
the equation reads \citep{Hassani:2021tdd}
\begin{align}
\label{PDE_1p1}
 \partial_{\tau}^2\pi  =  \left(\nabla \pi\right)^2 ,
\end{align}
and we consider the initial conditions 
\begin{eqnarray}
 \label{IC_1p1}
\pi(0 , x) &=& 0\,, \\ \nonumber
\partial_\tau \pi(\tau,x)\vert_{\tau=0} &=& {\alpha} \Psi(x) \,. 
\end{eqnarray}
First we show that in this case the minima and maxima of the scalar field $\pi$ do not move in space, 
a property that we also validate 
numerically, see Fig.\ \ref{Numerics_1D_blowup}. 
We then derive a PDE for the curvature of the scalar field, which at the extrema satisfies an ODE. We explicitly compute the (finite) time at which the curvature of minima becomes infinite.\footnote{In \cite{Hassani:2021tdd} we write the spatial dependence of $\pi$ near an extremum as a quadratic function of $x$, which provides a particular solution of the PDE.}

Let us define $D \equiv \nabla \pi$.
Taking the spatial derivative of the PDE~\eqref{PDE_1p1} results in a new equation for $D(\tau,x)$,
\begin{equation}
 \label{eq_k}
\partial_\tau^2 D = 2  D \nabla D
 \, .
\end{equation}
As $D(0,x) = 0$ according to Eq.~\eqref{IC_1p1}, it also implies that $ \dd_\tau^2 D(\tau,x)\vert_{\tau = 0} = 0$. On the other hand we have 
\be
 \dd_\tau D(\tau,x)\vert_{\tau=0} = {\alpha} \nabla \Psi(x)\,. \ee
It is then evident that for any points $x_s$ that are locations of extrema of $\Psi$, $D(\tau, x_s) = 0$ at all times, \ie\ these points are also extrema of $\pi$ (which remain fixed in position).

Taking a further spatial derivative of Eq.~\eqref{eq_k}, we obtain a PDE for the curvature of the scalar field,
\begin{align}
 \label{eq_curvature_app}
\partial_\tau^2 \kappa = 2  \kappa^2 + 2 D \nabla\kappa\, ,
\end{align}
where $\kappa(\tau,x) \equiv \nabla D(\tau,x)$ is the curvature.
In general this equation is not closed, as we need $\nabla\kappa(\tau,x)$ to solve the equation and this term is obtained through a higher-order derivative PDE (by taking another spatial derivative of the equation). However, 
for the extremal points $x_s$ where $D(\tau,x_s) = 0$ we can close the equation as the second term vanishes,
\begin{align}
 \label{closed_eq_app}
\partial_\tau^2 \kappa (\tau,x_s) = 2  \kappa(\tau,x_s)^2 
 \, .
\end{align}
This is an ODE for the evolution of the scalar field curvature at the extrema. The initial conditions for $\kappa(\tau,x_s)$ are obtained using the initial profile of $\pi$ and $\dd_{\tau} \pi$ in Eq.~\eqref{IC_1p1}, 
\begin{eqnarray}
\pi(0 , x) = 0 &\longrightarrow& \kappa (0,x_s) = 0\,, \\ \nonumber
 \nabla^2 \partial_\tau \pi(\tau,x)\vert_{\tau=0} = {\alpha} \nabla^2 \Psi(x) &\longrightarrow& \partial_\tau \kappa(\tau,x_s)\vert_{\tau=0} = {\alpha} \nabla^2 \Psi(x_s)\,.
\end{eqnarray}
With these initial conditions a first integral of Eq.~\eqref{closed_eq_app} yields 
\begin{equation}
\left(\partial_\tau\kappa\right)^2 = \left({\alpha} \nabla^2 \Psi\right)^2 
+ \frac{4}3 \kappa^3 \,,
\end{equation}
where we have dropped the argument $x_s$ for brevity.
 Considering the point $x_s$ being a minimum, \ie\ $\partial_\tau \kappa > 0$ and integrating the previous equation from $\tau=0$ to the blow-up time $\tau_b$ such that at this time the curvature goes to infinity, we obtain
\be
\int_{0}^{\infty} \frac{d \kappa}{\sqrt{\left({\alpha} \nabla^2 \Psi\right)^2 + \frac{4}3 \kappa^3}} = \int_0^{\tau_b} d\tau = \tau_b
\ee
Changing the integration variable from $\kappa$ to $s$ where $s^{3}={}^4\!/_{\!3}\,\kappa^3/\left({\alpha} \nabla^2 \Psi\right)^2$ 
we find 
\begin{equation}
\tau_{b}=\left(\frac{3}{4 {\alpha} \nabla^2 \Psi}\right)^{\frac{1}{3}}
\int_{0}^{\infty} \frac{d s}{\sqrt{1+s^{3}}} = \frac{2 \Gamma\left(\frac{1}{3}\right) \Gamma\left(\frac{7}{6}\right)}{\sqrt{\pi}}  \Big(\frac{3}{4 {\alpha} \nabla^2 \Psi} \Big)^{\frac13}\,,
\label{generic_blowup}
\end{equation}
or
\be
\label{app_blowup_time}
\tau_b = 2.5479...  \; \left({\alpha} \nabla^2 \Psi(x_s) \right)^{-\frac13}\,.
\ee
To sum up, the minima blow up 
in a finite time given by Eq.~\eqref{app_blowup_time} which depends on the initial curvature of the potential $\Psi$ at the minimum. It is worth mentioning that if $x_s$ is a maximum, it can also become unstable depending on the initial value of $\partial_\tau\kappa$.
Based on the ODE \eqref{closed_eq_app}, 
$\dd_{\tau}^2\kappa$ is always positive as it is sourced by $\kappa^2$. So we roughly expect that the curvature of maxima 
starts to increase and 
eventually becomes flat and switches sign
after which it blows up in finite time given by Eq.~\eqref{generic_blowup}. In \cite{JPE_Hassani} the dynamics of maxima and minima, especially when $c_s^2 \neq 0$ is studied in more detail.

\begin{figure}[tb]%
    \centering
    \hspace*{-1cm}  
   {{\includegraphics[scale=0.4]{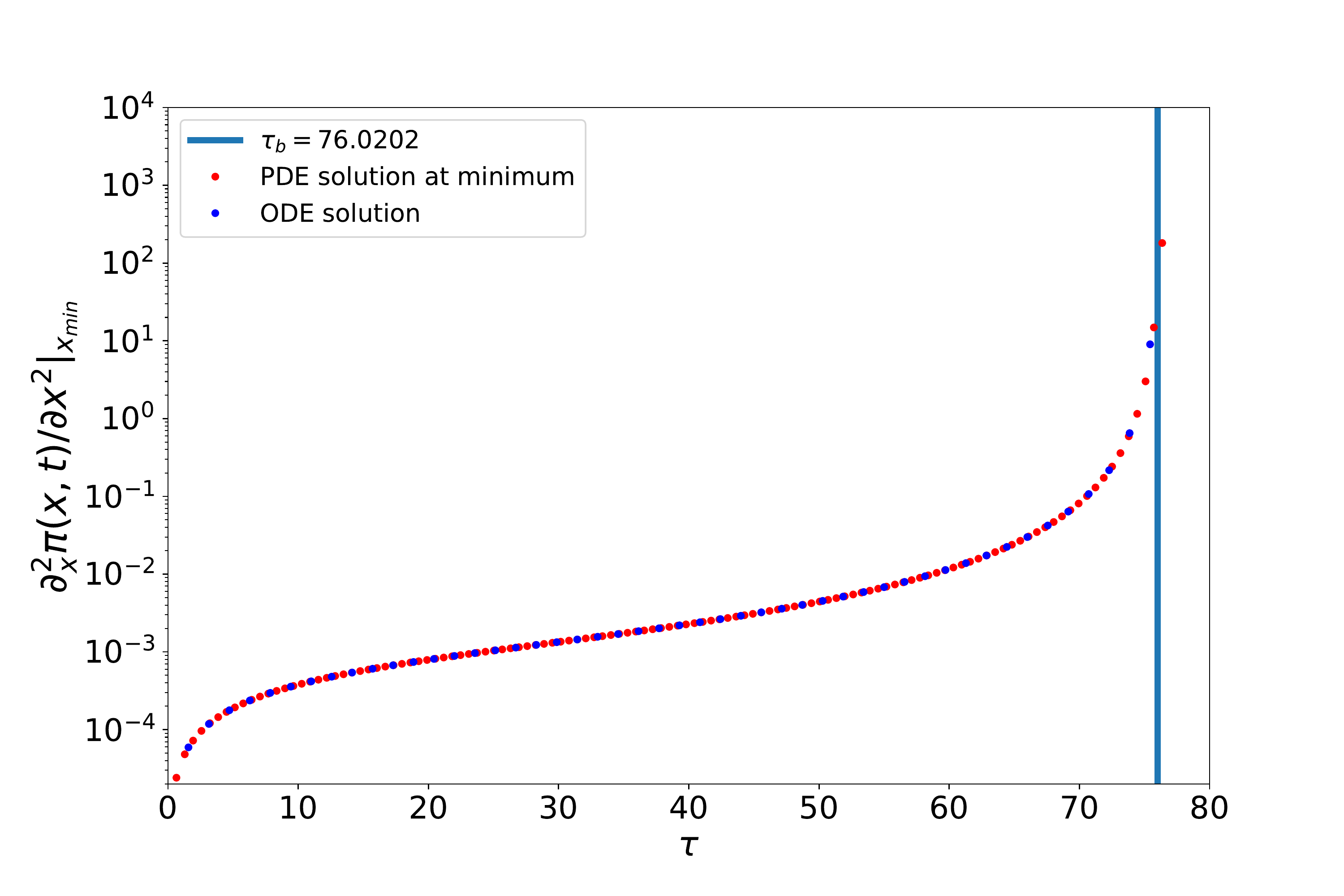} }}%
        \caption{We show the time evolution of the curvature at the minimum. The solution of the ODE \eqref{closed_eq_app}, the analytical solution in Eq.~\eqref{generic_blowup} and the curvature evolution obtained from the solution of the PDE \eqref{PDE_1p1} are consistent and all blow up at the expected time.}
    \label{fig:extrema_blowup}%
\end{figure}

In Fig.~\ref{Numerics_1D_blowup} we show the numerical results for the scalar field $\pi(\tau,x)$ profile and its first and second time derivative in the top panel and its spatial derivatives in the bottom panel. The scalar field $\pi(\tau,x)$ and its derivatives are obtained by solving the PDE~\eqref{eq_oneD} numerically on a lattice in 1+1 D assuming $c_s^2 = 0$, periodic boundary conditions and $\Psi(x) = \cos (4 \pi x/L) \;$\footnote{The reason for considering
a periodic function like the cosine instead of $x^2$ is that periodic boundary conditions are easy to implement and do not require additional assumptions.}. Here $N_{\rm grid} = 2048$  is the number of points and we choose the units such that $\alpha = 1$ and $dx = 1$ is the distance between the points on the 1D lattice. Hence %so that
$L=N_{\rm grid}=2048$ and as a result $\nabla^2 \Psi(x_s) = 3.765 \times 10^{-5} $ which based on Eq.~\eqref{app_blowup_time} blows up at $\tau_b = 76.02$.
The curvature of the maxima and minima increases with time so that the maxima become flatter while the minima become sharper and eventually blow up at a finite time given by Eq.~\eqref{generic_blowup}. Moreover, paying attention to $\nabla \pi$ in the middle part of the bottom panel of the figure, one realises that this function shares similar behaviour with caustic singularities \cite{ArkaniHamed:2005gu, Babichev:2016hys, Babichev:2017lrx}  which comes from the fact that to leading order according to Eq.~\eqref{SEfullfluids} the velocity $v_x$ is $\nabla \pi$ in 1+1 D and the maxima and minima (of the velocity) travel towards each other to form a caustic in finite time. To verify our analytical results, we compare the blow-up time obtained from the solution of the ODE \eqref{closed_eq_app}  with the numerical solution from the PDE at the minimum point in Fig.~\ref{fig:extrema_blowup}. According to the figure our theoretical solution and the numerical results agree very well.
Solving the PDE~\eqref{eq_oneD} for large $c_s^2/\alpha$ changes the behaviour of the system from a divergent to a stable one. For example for the limit $\alpha \to 0$ and $c_s^2 \neq 0$ we have a stable wave solution.
 
Similar to the case of spherical symmetry, for $c_s^2>0$ one can use a Hopf-Cole transformation to obtain a particular exact solution of Eq.~\eqref{eq_oneD},
\begin{equation}
    \pi(\tau,x) = \frac{c_s^2}{\alpha} \left(\ln\cosh \frac{C x}{c_s} + \frac{1}{2} C^2 \tau^2\right)\,,
\end{equation}
where $C$ is an arbitrary constant.
In this solution the spatial profile is again constant, as in the spherically symmetric case. However, we find numerically that the spatial profile evolves for more generic initial conditions. But
 the existence of non-singular solutions for $c_s^2 > 0$ indicates that the limit of small speed of sound needs to be taken with great care, and that further investigations into the physical behaviour in this limit may be warranted\footnote{
 In a detailed study in \cite{JPE_Hassani} this limit will be discussed for non-zero initial conditions with compact support.}.

%%%%%%%%%%%%%%%%%
%%%Section  : Conclusions
%%%%%%%%%%%%%%%%%
\section{Conclusions and discussion}
Our results show that the EFT description of $k$-essence DE breaks down for part of the parameter space due to a non-linear instability triggered by a term $\propto (\vec\nabla \pi)^2$ in the equations of motion. Based on our numerical analysis using realistic cosmological simulations we see that this instability happens only when the speed of sound $c_s$ is small such that the stabilizing linear term $c_s^2 \nabla^2 \pi$ is suppressed. To gain some analytical insight we investigated the PDE in simplified setups. First, we considered a spherically symmetric scenario and showed that the non-linear term $(\vec\nabla \pi)^2$ does indeed lead to a blow-up and a similar behaviour to what we saw in the three-dimensional simulations. We specifically showed that the relation between the blow-up redshift and the initial density of the potential well is similar to what we find in cosmological simulations which implies that the instability is found correctly. We further, in connection with a mathematical discussion, studied a simplified PDE considering only the two important terms. We showed that the system, similar to our simulation results, is unstable for vanishing speed of sound. Moreover, we find numerically that the stability is gained for large values of speed of sound. We derived an ODE for the curvature of the minima and showed that the curvature goes to infinity in a finite time. We compare the numerical 1+1 D solution of the PDE at the minimum point, with the ODE and the blow-up time prediction and find consistent results.

The non-linear instability we found does not appear in the linearized theory, where the evolution is stable for all values of the speed of sound. The presence of such an instability shrinks the 
$w-c_s^2$ parameter space for the healthy $k$-essence type theories when treated within the EFT framework. Moreover, similar terms appear in non-linear parametrisations of the EFT framework in the context of more general theories which can therefore suffer from related instabilities. As a result these theories have to be considered more carefully, particularly when $(\vec\nabla \pi)^2$ appears in the scalar field equation of motion. 

The breakdown of the EFT approach can be either due to the EFT truncation order where higher-order corrections can remove the instability, or it can be a hint of the full theory breakdown. In the case of the latter, it also leads to the breakdown of the weak-field approximation and requires a more careful analysis to decide whether coupling the scalar field to gravity could hide the singularities behind (black hole) horizons without a complete, global breakdown of the evolution. It is difficult to assess whether a tiny but non-vanishing $c_s^2$ is able to prevent a singularity, but even if this is the case the solutions will strongly depend on non-linearities and the truncation of the EFT is still rendered inconsistent.

%%%%%%%%%%%%%%%%%
%%%Acknowledgements  : 
%%%%%%%%%%%%%%%%%
\section*{Acknowledgements}
We would like to thank Emilio Bellini, Camille Bonvin, \O yvind Christiansen, Pierre Collet, Ruth Durrer, Jean-Pierre Eckmann, Ernst Hairer, Mona Jalilvand, David Mota, Sabir Ramazanov, Cornelius Rampf, Ignacy Sawicki, Edriss Titi, Filippo Vernizzi, Alexander Vikman, Hans Winther, Hatem Zaag and Miguel Zumalac\'arregui for many insightful discussions and valuable inputs during this project. 
\\
FH would like to especially express his gratitude to Jean-Pierre Eckmann for his invaluable support during this project and his comments about the manuscript.
This work was supported by a grant from the Swiss National Supercomputing Centre (CSCS) under project ID s1051. We acknowledge funding by the Swiss National Science Foundation.

%%%%%%%%%%%%%%%%%
%%%Appendices  : 
%%%%%%%%%%%%%%%%%

\appendix

%%%%%%%%%%%%%%%%%
%%%Appendix  : 
%%%%%%%%%%%%%%%%%
\section{Convergence tests \label{convergence_tests}}
\begin{figure}[tb]%
    \centering
    \hspace*{-1cm}  
   {{\includegraphics[scale=0.4]{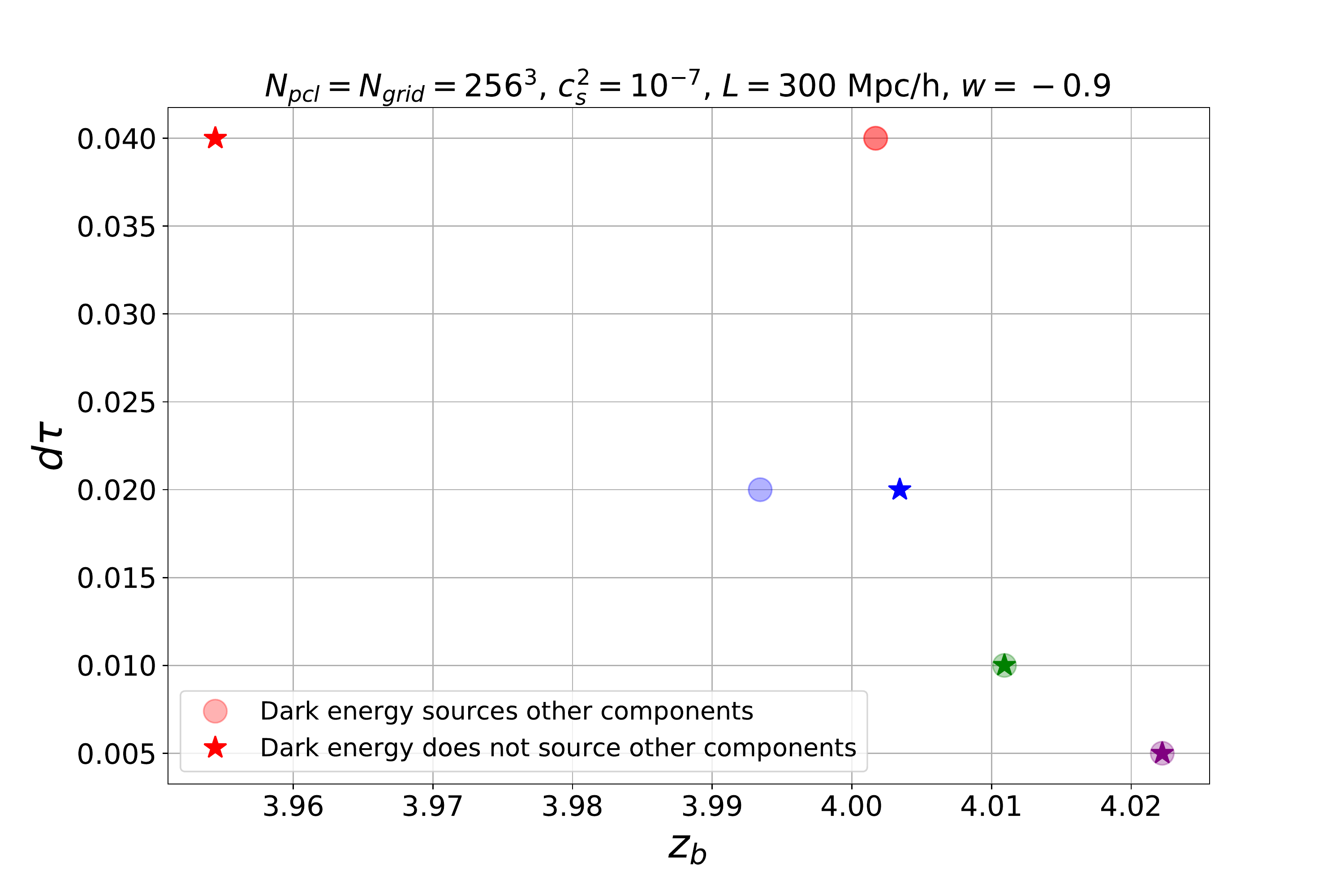} }}%
        \caption{The blow-up redshift for different time resolution (in units of the Hubble time) is plotted. The stars and circles represent the case when DE perturbations, respectively, source and do not source other components. As we increase the time precision, \ie\ decrease $d\tau$ in the  simulation, the blow-up time converges. Even for the lowest time precision there is no significant change in the blow-up redshift.} 
    \label{temporal_resolution}%
\end{figure}
In this appendix we discuss the tests performed to validate the results presented in Sec.~\ref{sec_3_numerics} for the realistic cosmological simulations. One important question we discuss here is %how much our results are robust and cosmology dependent, to answer this question we need to check 
the impact of precision parameters as well as the assumed cosmology on the blow-up phenomena. %As a result, in this appendix in the first part we check the impact of  temporal precision
In particular, we first check the robustness of the time integration of the $N$-body code, and then discuss extensively
how the blow-up redshift depends on the spatial resolution of the simulation. The latter is relevant
because increasing the resolution results in higher probability for deeper potential wells and changes the initial conditions for the scalar field accordingly. Finally we discuss how much the critical value of the speed of sound $c_s^2 \sim 10^{-4.7}$ depends on our cosmological assumptions and validate that the stability of the system is obtained for large values of $c_s^2$ even if we consider matter domination and go far beyond the present epoch. 
These tests, in addition to our simplified scenarios (spherical and planar symmetries) in Sec.~\ref{sec:one_d}, rule out the possibility of the instability being an artefact.

% Temporal
\subsection{Precision of the time integration\label{sec:timeprec}}
In this part we discuss the effect of the %time 
precision of the time integration on our results. In Fig.~\ref{temporal_resolution}, for a specific case where $N_\mathrm{pcl} = N_\mathrm{grid} = 256^3$, $L=300$ Mpc/$h$ and $c_s^2 = 10^{-7}$, we show the sensitivity of the blow-up redshift to the time resolution of the simulation. For each %time 
precision setting we consider two simulations, one where DE perturbations source other components 
(stars), and one where this is not the case, \ie\ DE is a spectator field (circles).
Based on these results,
even for the largest $d\tau$ (the lowest precision considered) the blow-up redshift does not change significantly.
This 
also shows that the blow-up redshift (for high enough time resolution) does not depend on gravity being sourced by the DE 
component.  In summary, our test indicates that the time precision considered in our cosmological simulations is sufficient 
to resolve the blow-up phenomenon.

% spatial
\subsection{Spatial resolution \label{app_spatial}}

\begin{figure}[tb]%
    \centering
    \hspace*{-1cm}  
   {{\includegraphics[scale=0.4]{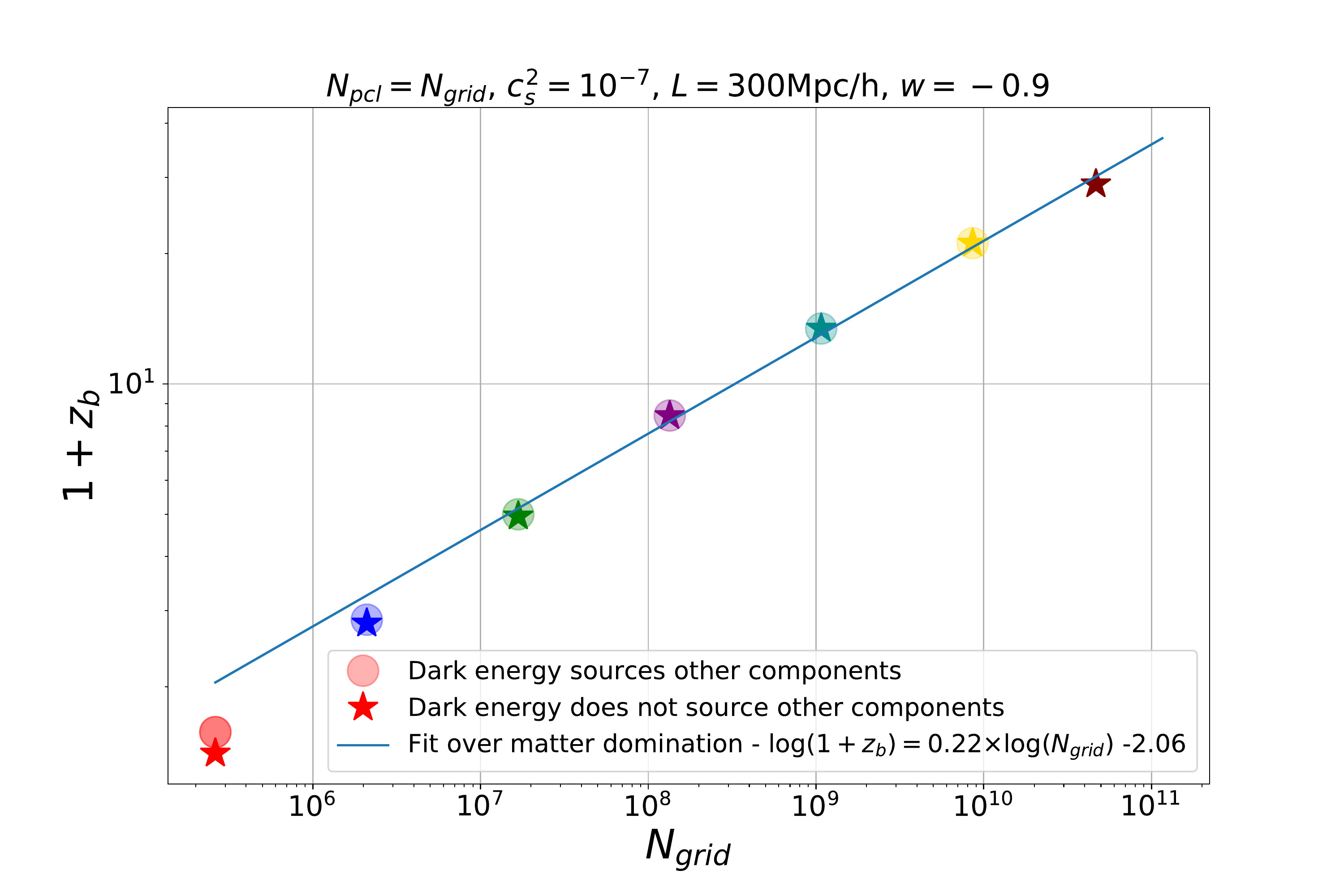} }}%
        \caption{The blow-up redshift for 
        a fixed box size and different numbers of grid points is shown. The fit over the data where the blow-up happens in matter domination results in the relation $1+z_b \sim N_\mathrm{grid}^{0.22}$. The circles/stars represent the case where the DE component does/does not source gravity.}
    \label{spatial_resolution_v2}%
\end{figure}

\begin{figure}[tb]%
    \centering
    \hspace*{-1cm}  
   {{\includegraphics[scale=0.4]{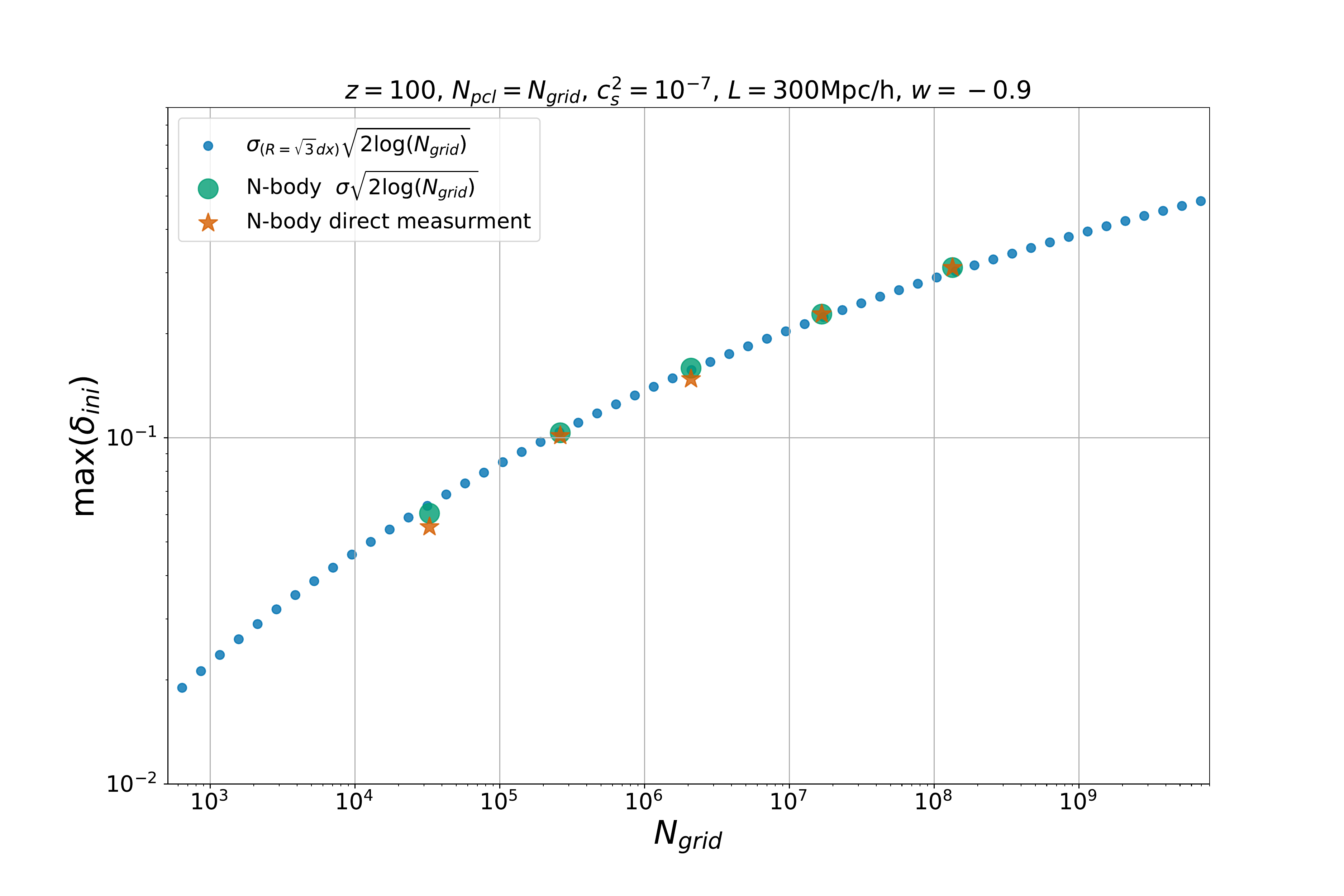} }}%
        \caption{The maximum of the initial density contrast at $z=100$ as a function of the number of grid points is shown. The orange stars correspond to the direct measurement of $\max(\delta_\mathrm{ini})$ on the
        initial snapshot of the simulation. The green and blue circles represent $\max(\delta_\mathrm{ini})$ computed using the formula \eqref{App:mean_gumbel} for the mean of the Gumbel distribution. The standard deviation used for the green circles is measured
        directly from the snapshots while
        for the blue circles we use 
        Eq.~\eqref{app:sigma_R} and using the linear power spectrum from linear Boltzmann code CLASS \cite{Blas:2011rf}.}
    \label{Fig_Gumbel}%
\end{figure}

In this subsection we discuss the dependence of the blow-up redshift on the spatial resolution of our simulations. Contrary to the time precision, the spatial resolution is not only a precision parameter of the equation, but also a parameter which changes the initial conditions. Higher-resolution simulations probe smaller scales where the amplitude of perturbations is larger, and as a result the scalar field evolution is sourced by deeper potential wells. In Fig.~\ref{density_z_b} we show the blow-up redshift for different maximum values of the initial density, $\max(\delta_\mathrm{ini})$, where we obtain a linear relation between the two, \ie\ $1+z_b \propto \max(\delta_\mathrm{ini})$ in the matter-dominated era. In Fig.~\ref{spatial_resolution_v2} we show the measurement of the blow-up redshift for different spatial resolutions. Our fit over the data in matter domination shows $1+z_b \propto N_{\rm grid}^{0.22}$ where $N_{\rm grid}$ is the total number of grid points in 3D.
Our results in Fig.~\ref{density_z_b} and Fig.~\ref{spatial_resolution_v2}, considering only 
the data in matter domination then suggest
\be
\max(\delta_{ini}) \sim N_{\rm grid}^{0.22}\,. 
\ee 
In cosmological $N$-body simulations one can estimate the maximum of the initial matter density contrast, $\max(\delta_\mathrm{ini}(x))$, without measuring it directly through the snapshots
by invoking
the Fisher-Tippett-Gnedenko extreme value theorem \cite{book}. The extreme value theorem describes the distribution of extrema in a similar fashion to how the central limit theorem concerns the behaviour of averages. It states that for a large number $N$ draws from a normal distribution\footnote{The extreme value theorem is far more general and can be applied to large samples drawn from quite general distributions, as is the case for the central limit theorem.} with average $\mu$ and standard deviation $\sigma$, the maximum of the sample follows a Gumbel distribution with a mean approximated as
\be \label{App:mean_gumbel}
\langle \max  \rangle \propto \mu + \sigma \sqrt{2 \ln N}\,.
\ee

In a cosmological simulation  
we may assume that at early times the density contrast follows a normal distribution $\mathcal{N}(\mu,\sigma)$ with vanishing mean, $\mu = 0$, and a standard deviation $\sigma_R$ that depends on the resolution of the simulation, corresponding to a smoothing scale $R$, and is given by
\begin{equation} \label{app:sigma_R}
\sigma_{R}^2=\frac{1}{2 \pi^{2}}\int_{0}^{\infty} P(k) W(k, R)^{2} k^{2} dk \, .
\end{equation}
Here $W(k, R)$ is the Fourier transform of a window function with radius $R$, and $P(k)$ is the power spectrum.
We assume a top-hat window function with radius $R$, with Fourier transform,
\begin{equation}
W(k, R)=\frac{3[\sin (k R)-k R \cos (k R)]}{(k R)^{3}} \, .
\end{equation}
As the standard deviation of the density contrast depends on the resolution of a simulation through $R \sim dx \propto N_\mathrm{grid}^{-1/3}$, we see that, for fixed physical size of the simulation cube, larger $N_\mathrm{grid}$ will lead to smaller $R$ and thus
to higher wavenumbers contributing in the integral, resulting in a larger variance. Based on our numerical measurements, for the choice $R = \sqrt{3} dx$ the result of the integral \eqref{app:sigma_R} agrees with our numerical measurements of the variance. In Fig.~\ref{Fig_Gumbel} we show the numerical results for $\max(\delta_\mathrm{ini})$ computed in three different ways. The orange stars represent the direct measurement of  $\max(\delta_\mathrm{ini})$ in the initial snapshot of the $N$-body simulation for a given $N_\mathrm{grid}$, the green and blue circles represent the results when we use the Gumbel distribution \eqref{App:mean_gumbel} to compute  $\langle\max(\delta_\mathrm{ini}) \rangle$  where in the latter case we use $\sigma_{R = \sqrt3 dx}$ computed through the integral \eqref{app:sigma_R} whereas for the former case we use $\sigma$ as measured directly from the snapshot of the $N$-body simulation. In both cases we assume $N = N_\mathrm{grid}$, even though the draws are not entirely independent. The figure shows excellent
agreement between different approaches for computing $\max(\delta_\mathrm{ini})$.
As a result, we can estimate the maximum of the initial density in an $N$-body simulation using Eq.~\eqref{App:mean_gumbel} where the variance is computed by Eq.~\eqref{app:sigma_R}. 

\subsection{Cosmology dependence of $c_s^*$ \label{cs_app}}
\begin{figure}[tb!]%
    \centering
    \hspace*{-1cm}  
   {{\includegraphics[scale=0.4]{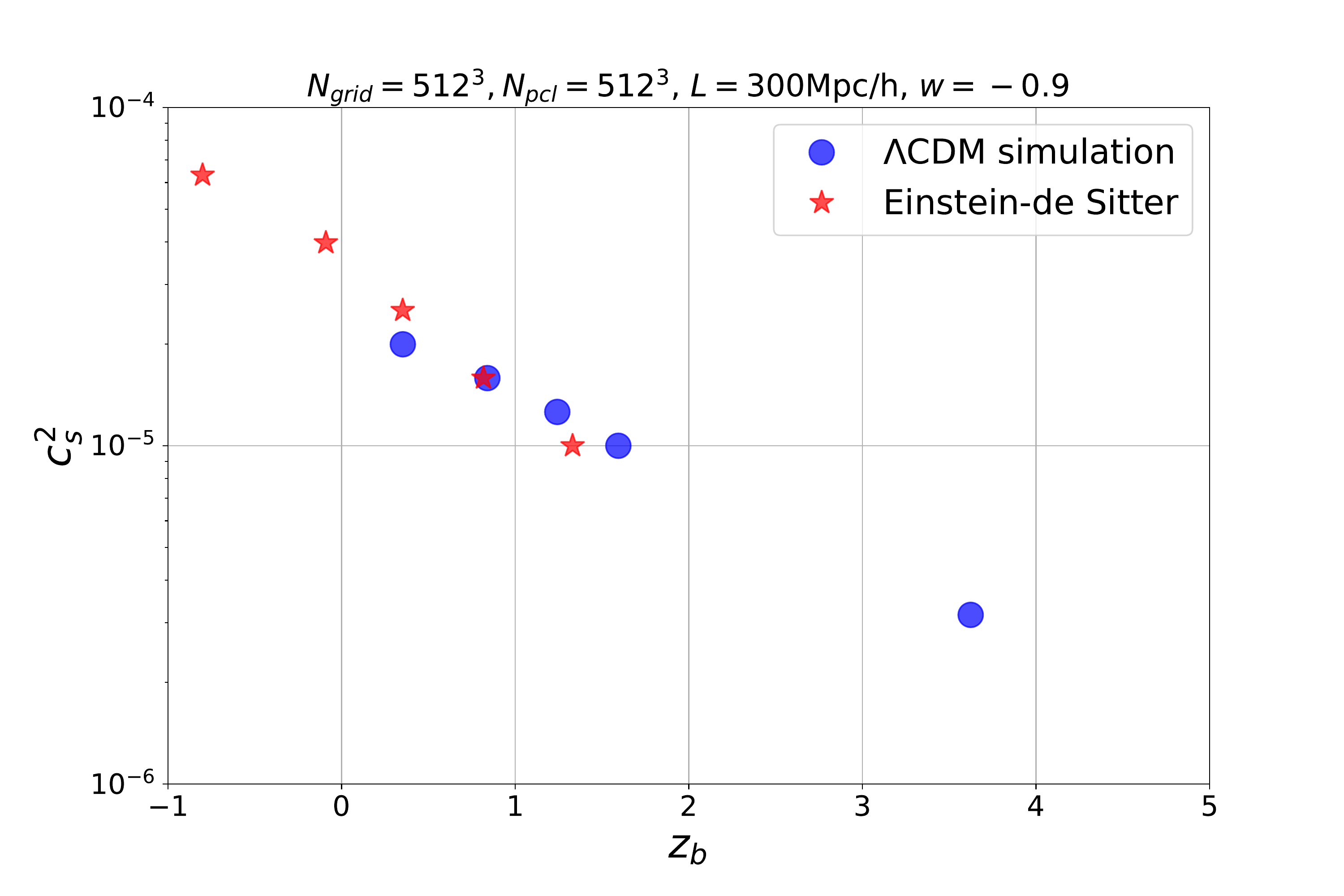} }}%
        \caption{The blow-up redshift for different speeds of sound is shown. The blue circles represent the $\Lambda$CDM scenario, while the red stars show the matter domination case \ie $\Omega_m =   1$  where we choose $h=0.3775$ such that $w_m / h^2 =1$. In the $\Lambda$CDM case letting the simulation run to the future (negative redshifts) does not decrease the blow-up %sound speed 
        threshold for the speed of sound. This happens because in the DE dominated era the potential wells decay and help to stabilise the system. However, in the Einstein--de~Sitter %matter domination 
        scenario, going to negative redshifts can increase the 
        threshold to $ c_s^2  \approx 10^{-4.35}$.}
    \label{inital_redshift_test}%
\end{figure} 
In Sec.~\ref{Sec:1_eq} 
we showed that there is a critical speed of sound $c_s^*$ such that the system remains stable for $c_s > c_s^*$. As the blow-up redshift generally decreases with increasing $c_s$, there is the possibility that the critical value is connected to the onset of DE domination
where the stability of the system is guaranteed due to decay of potential wells. In this subsection we  rule out this possibility by considering simulations in Einstein--de~Sitter cosmology, \ie\ where matter always dominates. In order to simulate a Universe with $\Omega_m = 1$ while being close to the $\Lambda$CDM Universe at early times (including the radiation era) we choose a different Hubble parameter $h=0.3775$ such that $\omega_m/h^2=1$. 
In Fig.~\ref{inital_redshift_test} we show the results for $c_s^2$ when we have matter domination compared to the $\Lambda$CDM scenario. In the $\Lambda$CDM case, due to the late-time DE domination, the PDE does not blow up in the future as the potential wells decay and we obtain the critical value of $c_s^2 \approx 10^{-4.7}$. On the other hand, in Einstein--de~Sitter 
the system could still blow up in the future. However, there is still a critical value for the speed of sound $c_s^2 \approx 10^{-4.35}$ where for larger values of the speed of sound the system remains stable and does not blow up even in the far future. This observation is in agreement with the claim that for large $c_s$ the stability of the system is restored due to the pressure term in the equation, and is not
an effect of late-time DE domination.

%%%%%%%%%%%%%%%%
%%%%%%%%%%%%%%%%
\bibliographystyle{JHEP}
 %\clearpage
\bibliography{bibliography}
%%%%%%%% End of the .bbl content 

\end{document}